\documentclass[twocolumn]{aastex631}
\pdfoutput=1 
\usepackage{amsmath,amssymb,amstext}
\usepackage{gensymb}
\usepackage{url}
\usepackage{comment}
\usepackage{tabularx}
\usepackage{array}
\usepackage[normalem]{ulem}
\usepackage{enumitem}

\def\lsim{\;\rlap{\lower 2.5pt
   \hbox{$\sim$}}\raise 1.5pt\hbox{$<$}\;}
\def\gsim{\;\rlap{\lower 2.5pt
 \hbox{$\sim$}}\raise 1.5pt\hbox{$>$}\;}

\usepackage{appendix}

\submitjournal{ApJ}

\shorttitle{Reacceleration of Streaming Cosmic Rays}
\shortauthors{Bustard \& Oh}

\begin{document}

\title{Turbulent Reacceleration of Streaming Cosmic Rays}

\correspondingauthor{Chad Bustard}
\email{bustard@ucsb.edu}
\author[0000-0002-8366-2143]{Chad Bustard}
\affil{Kavli Institute for Theoretical Physics, University of California - Santa Barbara, Kohn Hall, Santa Barbara, CA 93107, USA}

\author{S. Peng Oh}
\affil{Department of Physics, University of California - Santa Barbara, Broida Hall, Santa Barbara, CA 93106, USA}

\begin{abstract}
Subsonic, compressive turbulence transfers energy to cosmic rays (CRs), a process known as non-resonant reacceleration. It is often invoked to explain observed ratios of primary to secondary CRs at $\sim \rm GeV$ energies, assuming wholly diffusive CR transport. However, such estimates ignore the impact of CR self-confinement and streaming. We study these issues in stirring box magnetohydrodynamic (MHD) simulations using Athena++, with field-aligned diffusive and streaming CR transport. For diffusion only, we find CR reacceleration rates in good agreement with analytic predictions. When streaming is included, reacceleration rates depend on plasma $\beta$. Due to streaming-modified phase shifts between CR and gas variables, they are slower than canonical reacceleration rates in low-$\beta$ environments like the interstellar medium (ISM) but remain unchanged in high-$\beta$ environments like the intracluster medium (ICM). We also quantify the streaming energy loss rate in our simulations. For sub-Alfv\'{e}nic turbulence, it is resolution-dependent (hence unconverged in large scale simulations) and heavily suppressed compared to the isotropic loss rate $v_{A} \cdot \nabla P_{\rm CR} / P_{\rm CR} \sim v_{A}/L_{0}$, due to misalignment between the mean field and isotropic CR gradients. Unlike acceleration efficiencies, CR losses are almost independent of magnetic field strength over $\beta \sim 1-100$ and are, therefore, not the primary factor behind lower acceleration rates when streaming is included. While this paper is primarily concerned with how turbulence affects CRs, in a follow-up paper, we consider how CRs affect turbulence by diverting energy from the MHD cascade, altering the pathway to gas heating and steepening the turbulent spectrum.

\end{abstract}

\section{Introduction}

Cosmic rays (CRs) are an important non-thermal component of galaxies and their surroundings. In the Milky Way interstellar medium (ISM), their collective energy density is comparable to that in thermal gas, magnetic fields, and turbulence \citep{Boulares1990}, and they are believed to play a significant role in ionizing molecular clouds (e.g., \citealt{dalgarno06}), driving galactic outflows \citep{Ipavich1975, breitschwerdtwinds1991, SalemCROutflows2014, Ruszkowski2017, Buck2020, Bustard2020, Hopkins2020}, and impacting pressure balance, stability, and fragmentation of the ISM \citep{Parker1966, Heintz2020}. Their long residence times (compared to the light crossing time) in the disk, inferred from ratios of their spallation products, suggest that CRs are frequently scattered by magnetic perturbations, forcing them to undergo a random walk along magnetic field lines.

Such frequent scatterings drive the galactic CR distribution to almost perfect isotropy, making it difficult to connect CRs back to their sources; however, detections of their secondary by-products give us clues to their origins and transport throughout the Galaxy. The ``standard paradigm", implemented in most phenomenological propagation models (e.g. \citealt{Strong1998, Dragon2}), is one in which galactic CRs are predominantly created by diffusive shock acceleration (DSA) at supernova remnant shock fronts, followed by energy-dependent, diffusive propagation. This diffusive picture is physically motivated for high-energy CRs above a few hundred GeV, for which we believe CRs are confined by scattering off hydromagnetic waves created by an external turbulent cascade \citep{Chandran2000, Yan2004}. 

For $E \lessapprox 300$ GeV, however, the dominant confinement mode appears to be self-confinement, where CRs pitch angle scatter off Alfv\'{e}n waves that the CRs generate themselves through a resonant streaming instability \citep{Wentzel1968, Kulsrud1969}. In this energy range, the resulting transport is a mixture of field-aligned streaming down the local CR pressure gradient at the Alfv\'{e}n speed $v_{A,i} = B/\sqrt{4\pi\rho_{i}}$ and field-aligned diffusion that arises from wave damping (Section \ref{sec:analytic_streaming}) and the subsequently reduced pitch angle scattering rate (see recent reviews by e.g. \citealt{Zweibel2013, ZweibelReview2017, AmatoReview2018}). Self-confinement is not only supported by a growing number of theoretical studies but also by a break in the primary spectrum at 300 GeV, which aligns with the theoretically expected transition from self-confinement to extrinsic turbulence \citep{Blasi2012, Aguilar2015, Evoli2018, AmatoReview2018, Evoli2019, Kempski2021}. While this model is increasingly implemented in both galaxy evolution and CR propagation models, there are still many unknown outcomes of diffusive vs streaming transport, both for the interpretation of observables and for our understanding of CR influence on galaxy evolution. The one we concern ourselves with in this paper is turbulent reacceleration.

CRs in a turbulent background can gain energy from both resonant and non-resonant interactions through second order Fermi acceleration \citep{Brunetti2007, Lynn2013}. The particle mean free path $\lambda_{\rm mfp}$ relative to the lengthscale $l$ of the turbulent eddy it interacts with determines the relevant regime. In the low scattering rate limit ($\lambda_{\rm mfp} >> l)$, the key interactions are resonances between particles and magnetic compressions due to fast or slow magnetosonic waves\footnote{The strong anisotropy of the Alfv\'{e}nic cascade at small scales means that scattering is highly inefficient \citep{Chandran2004,Yan2004}.}. Although strong damping generally precludes the turbulent cascade from extending down to the CR gyro-scale, CRs with parallel velocities similar to the wave speed can extract energy from it, a process known as transit-time damping (similar to Landau damping). Thus, weakly scattered CRs undergo {\it resonant} acceleration. 

However, we have argued that low-energy $E < 300$ GeV CRs in our Galaxy are self-confined, with small mean free paths. This is particularly true of the $\sim$GeV CRs, which carry most of the energy and have $\lambda_{\rm mfp} \sim $pc scale mean free paths in $ \sim \mu$G magnetic fields. Since they are strongly scattered ($\lambda_{\rm mfp} << l$), the strong coupling with the gas means that they undergo compression and rarefaction with the thermal fluid. If the CRs were completely locked to the fluid, these constitute adiabatic reversible processes, and CRs experience no net energy gain. However, CR diffusion out of overdense regions breaks this symmetry, so that there is net energy transfer from the gas to the CRs \citep{Ptuskin1988}. Heuristically, the CRs gain energy during compression, but diffuse out without giving this energy back. This {\it non-resonant} reacceleration is the focus of this paper. In Galactic propagation models, reacceleration is typically included as a default, but it is in increasing tension with data, including both secondary-to-primary ratios as a function of energy  \citep{StrongReview2007, GabiciReview2019}, and a growing wealth of multiwavelength data, specifically synchrotron measurements \citep{Trotta2011, DiBernardo2013, Orlando2013, GabiciReview2019}.

The primary goal of this paper is to evaluate the effect of Alfv\'{e}nic CR streaming, which is characteristic of all self-confined CRs. Since CRs only stream out of local CR maxima, CR streaming also breaks the symmetry between compression and rarefaction, but its effects are far less known. The impact of CR streaming on turbulent reacceleration has only recently been discussed analytically \citep{Hopkins2021_reacc} and never been quantitatively studied in MHD simulations. We find that CR streaming reduces the efficiency of turbulent reacceleration in low $\beta$ environments, potentially explaining why significant reacceleration appears to be disfavored in models. Somewhat counterintuitively, though, CR streaming losses are {\it not} responsible for lower acceleration rates: we don't find a correlation between CR streaming loss rates and reaceleration rates. Interestingly, loss rates are quite different from isotropic loss rates $v_{A} \cdot \nabla P_{\rm CR} / P_{\rm CR} \sim v_{A}/L_{0}$, due to misalignment between the mean field and isotropic CR gradients, and this effect is resolution dependent and may be incorrectly captured in low-resolution galaxy evolution simulations.

Our paper is outlined as follows: First, we provide background on turbulent reacceleration and present the previously-derived growth rate for purely diffusive CRs subject to long-wavelength, subsonic, isothermal turbulence (\S \ref{sec:background}). We then extend this treatment to include additional self-confinement (streaming) terms, and we derive a simple modification to the canonical reacceleration rate of purely diffusive CRs. In \S \ref{sec:numerics}, we numerically validate these growth rates using a two-fluid CR implementation in the Athena++ code, and in \S \ref{sec:heating} we quantify the efficiency of streaming energy loss in turbulence with varying plasma $\beta$. In \S \ref{sec:discussion}, we discuss the implications of these results for studies of the ISM, intracluster medium (ICM), and circumgalactic medium (CGM), aimed at both the galaxy evolution and CR propagation communities. We conclude in \S \ref{sec:conclusions}. 

While in this paper we focus on how turbulence affects CRs, in a follow-up paper (Bustard and Oh, in prep), we study the back-reaction of CRs on turbulence. This has largely been neglected, although it is clear that CRs are absorbing energy from the flow. We study this both analytically and in exploratory simulations. We find that in many physically plausible scenarios relevant to the ISM and CGM (where CR energy densities are expected to be significant), CRs can absorb a large fraction or even most of the turbulent driving energy at large scales, significantly steepening the turbulent energy spectrum or even wiping out small scale compressive motions.

\section{Background and Analytic Arguments}
\label{sec:background}

Following \cite{Ptuskin1988}, let's define three scales: $L_0$ is the outer eddy scale set by large-scale driving motions; $L_1$ is some smaller ``cut-off" scale, which could be a viscous or damping scale, though here we will associate it with the width of a shock front in the medium\footnote{Even in subsonic turbulence, non-linear steepening will still produce weak shocks.}; and $L_{\rm turb} \in [L_1, L_0]$ represents a general eddy scale within the turbulent cascade. \cite{Ptuskin1988} derived the non-resonant acceleration rate from an ensemble of random acoustic waves and weak shocks propagating in compressive, subsonic\footnote{For transonic or supersonic turbulence, strong shocks can additionally accelerate CRs through the first order Fermi mechanism, but then the resulting acceleration rate and CR spectrum are qualitatively different from second order Fermi acceleration.} turbulence. We will quickly summarize those results in two illuminating limits (see also e.g. \citealt{Lynn2013}), first assuming as in \cite{Ptuskin1988} that CR transport is purely diffusive, before considering the effects of streaming.  

\subsection{Non-Resonant Reacceleration of Purely Diffusive Cosmic Rays}
\label{sec:analytic_purediff}

Let us first consider acceleration operating on a single scale, the turbulent outer scale $L_0$, with turbulent velocity $v$ and CR diffusion coefficient $\kappa$. An important insight from \cite{Ptuskin1988} is that in subsonic turbulence, the lifetime of a compression in subsonic turbulence is not the eddy turnover time $\tau_{\rm eddy} \sim L_0/v$ but instead the sound (or compressive wave) crossing time $\tau_{sc} \sim L_0/v_{\rm ph}$, which is shorter than $\tau_{\rm eddy}$. Here, $v_{\rm ph} \sim (P_{\rm tot}/\rho)^{1/2} \sim [P_{\rm g} + P_{\rm B} + P_{\rm CR})/\rho]^{1/2}$ is the compressible wave phase velocity, and is simply given by the gas sound speed $c_s \sim (P_{\rm gas}/\rho)^{1/2}$ if $P_g \gg P_{\rm B}, P_{\rm CR}$. The two relevant timescales are therefore the sound crossing time $\tau_{sc} \sim L_0/v_{\rm ph}$, and the diffusion time $t_{\rm diff} \sim L_0^{2}/\kappa$ across an eddy. 

First, consider the case where $t_{\rm diff} \ll t_{\rm sc}$, i.e. $\kappa >> v_{ph}L_0$. 

In this regime, quickly diffusing particles see an effectively static velocity field, and the derivation of the momentum diffusion coefficient, $D_{pp}$, for a CR with momentum $p$ follows that of the standard second order Fermi argument: $D_{pp} \sim (\Delta p)^{2} / \tau_{\rm scatter} \sim p^{2} v^{2}/(c^{2} \tau_{\rm scatter}) \sim p^{2} v^{2} / \kappa$. The energy growth time, accurate to within a factor of a few, is then 
\begin{equation}
    t_{\rm grow} \sim \frac{p^{2}}{D_{pp}} \sim \frac{\kappa}{v^{2}}; \quad \kappa >> v_{ph}L_0
    \label{eqn1}
\end{equation}

In the opposite limit, for which $\kappa << v_{ph}L_0$, CRs advect with the fluid and behave quasi-adiabatically during each compression and rarefaction: $\delta p/p \sim \delta P_{\rm CR} / P_{\rm CR} \sim \delta \rho / \rho \sim v/v_{\rm ph}$. Through successive compressions and expansions in a turbulent eddy of size $L_0$, the CRs undergo a random walk in momentum space until they manage to diffuse out of the eddy on a timescale $L_0^{2}/\kappa$. In this case, $D_{pp} \sim (\delta p)^{2}/\tau_{\rm diff} \sim (p^{2} v^{2}/v_{\rm ph} ^2) (\kappa/L_0^{2})$. Within a factor of a few, the growth time in this limit is
\begin{equation}
    t_{\rm grow} \sim \frac{p^{2}}{D_{pp}} \sim \frac{v_{\rm ph}^{2} L_{\rm 0}^2}{v^{2} \kappa}; \quad \kappa << v_{\rm ph}L_0
    \label{eqn2}
\end{equation}
Consider $\kappa \sim v_{\rm ph} L_0$, i.e. the case of maximally efficient acceleration. In this case, both Equations \ref{eqn1} and \ref{eqn2} reduce to\footnote{The alert reader will notice that this minimum growth time (Equation \ref{eqn:tmin_grow}) coincides with the cascade time of compressible fast modes of scale $l$ in MHD turbulence, $\tau_{\rm cas} \sim l v_{\rm ph}/v_{\rm l}^2$ \citep{nazarenko11}. This arises because both Fermi II acceleration and fast mode cascade timescales can be understood as the outcome of stochastic random walks. In Fermi II acceleration, $t_{\rm grow} \sim \tau_{\rm scatter} (\Delta E/E)^{-2} \sim \tau_{\rm scatter} (v/c)^{-2} \sim v_{\rm ph} L/v^{2}$, assuming that $\tau_{\rm scatter} \sim \kappa/c^{2} \sim v_{\rm ph} L/c^2$. In fast modes, two wave packets of scale $l$ interact on a wave crossing time $\tau_{\rm ph} \sim l/v_{\rm ph} \ll \tau_{\rm NL} \sim l/v_l$, where the non-linear steepening time $\tau_{\rm NL} \sim l/v$ is set by the non-linear advection term $v \cdot \nabla v$ in the Euler equation. Thus, each interaction results in a small velocity change $\Delta v/v \sim \tau_{\rm ph}/\tau_{\rm NL} \sim v/v_{\rm ph}$. The cascade time is then set by the number of interactions required for $\Delta v \sim v$, i.e. $\tau_{\rm cas} \sim \tau_{\rm ph} (\Delta v/v)^{-2} \sim \tau_{\rm ph} (v_{\rm ph}/v)^{2} \sim v_{\rm ph} L/v^2$.} 
\begin{equation}
t_{\rm grow} \sim \frac{v_{\rm ph} L_{\rm 0}}{v^{2}}; \quad \kappa \sim v_{ph}L_0,
\label{eqn:tmin_grow} 
\end{equation}

Consider now the contribution of multiple eddies in a turbulent cascade. If we sum contributions from a variety of scales, the general expression for the momentum diffusion coefficient is \citep{Ptuskin1988}: 
\begin{equation} 
D_{\rm pp} = p^{2} \frac{2 \kappa}{9} \int_{0}^{\infty} dk \frac{k^2 W_{\rm 1D}(k)}{v_{\rm ph}^2 + \kappa^2 k^{2}}
\label{eq:Dpp_Wk} 
\end{equation} 
where $W_{\rm 1D}(k)$ is the 1D turbulent power spectrum, normalized such that: 
\begin{equation} 
v^{2} = \int_{0}^{\infty} dk \ W_{\rm 1D}(k) 
\end{equation}
What is an appropriate expression for $W_{\rm 1D}(k)$? Recall that for hydrodynamic turbulence, a standard Hodge–Helmholtz decomposition usually shows that the compressive component of the velocity field is Burgers-like ($W_{\rm 1D}(k) \propto k^{-2}$), while the solenoidal component is Kolmogorov-like ($W_{\rm 1D} (k) \propto k^{-5/3}$). The Burgers-like component does not reflect a genuine cascade, but rather the appearance of shocks that directly transfer power from large to small scales. Such shocks occur even in subsonic turbulence, due to non-linear steepening of waves. In this case, inserting $W_{\rm 1D} \propto k^{-2}$ into Equation \ref{eq:Dpp_Wk} (see Equation 27 of \cite{Ptuskin1988} for the full form of the power-law $W_{\rm 1D}$(k), including pre-factors), and using $t_{\rm grow} \sim p^{2}/D_{\rm pp}$ gives, in the regime $v_{ph} L_1 << \kappa << v_{ph} L_0$ \citep{Ptuskin1988}:
\begin{equation}
\begin{split}
    t_{\rm grow} \sim \frac{9}{2} \frac{v_{\rm ph} L_0}{v^{2}} \left({\rm tan^{-1}}\left(\frac{\kappa}{v_{\rm ph}L_1}\right) - {\rm tan^{-1}}\left(\frac{\kappa}{v_{\rm ph}L_0}\right) \right)^{-1} 
\end{split}
\label{eqn:growthPtuskin}
\end{equation}

At scales outside of the power-law spectrum, i.e. at scales larger than $L_0$ or smaller than the cut-off scale $L_1$, the reacceleration time reverts to Equations \ref{eqn1} and \ref{eqn2}, respectively. The minimum reacceleration time of Equation \ref{eqn:growthPtuskin} is $t_{\rm grow} \sim (9/2) v_{\rm ph}L_0/v^{2}$, motivating one to write  
\begin{equation}
    D_{pp} \sim \frac{2}{9}\frac{p^{2}v^{2}}{\kappa}
\end{equation}

This is very similar to the resonant reacceleration rate in balanced turbulence \citep{Skilling1975, Thornbury2014, ZweibelReview2017}: $D_{pp} \sim 1/9 (p^{2}v_{A}^{2}/\kappa)$, which is essentially what is implemented in Galprop \citep{Strong1998} and other CR propagation codes. The diffusion coefficient is commonly taken to be a power-law in particle rigidity $\kappa \propto R^{\delta}$, where the diffusion power-law index, $\delta$, is related to the wave spectrum power-law index. In that case, there are additional pre-factors related to the wave energy and power-law index.  

Note that for $W_{\rm 1D}(k) \propto k^{-2}$, the integrand in Equation \ref{eq:Dpp_Wk} is proportional to $1/(v_{\rm ph}^{2} + \kappa^{2} k^{2})$, i.e. it diminishes rapidly at high $k$. Unlike resonant acceleration, the smallest scales do not dominate acceleration. This is good news, as it implies that non-resonant acceleration can be captured in MHD simulations. However, there {\it is} still resolution dependence in the $\kappa \ll v_{ph} L_0$ regime (see Appendix): as resolution increases, we find a slow increase in acceleration rates in our simulations. In fact, as seen for the analytic $L_0/L_1 = 1000$ case in Fig \ref{fig:appendixFig}, Equation \ref{eqn:growthPtuskin} gives a growth time which is almost constant for $\kappa < L_0 v_{\rm ph}$, i.e. it scales much more weakly with $\kappa$ than $t_{\rm grow} \propto 1/\kappa$, as obtained in the single eddy approximation (Equation \ref{eqn2}). Heuristically, we can understand this as follows. For $\kappa < L_0 v_{\rm ph}$, the acceleration is maximized at some smaller scale $l$, where $\kappa \sim l v_{\rm ph}$. However, for Burgers turbulence, the velocity at scale $l$ is $v_l \propto l^{1/2}$, so that the minimum growth time $t_{\rm grow} \sim v_{\rm ph} l/v_{l}^{2}$ (when $\kappa \sim l v_{\rm ph}$) is independent of $l$. In other words, as long as there is sufficient dynamic range, acceleration in Burgers turbulence is self-similar, with eddies at some scale $l \sim \kappa/v_{\rm ph}$ providing the dominant contribution, and other eddies in the $\kappa \ll v_{\rm ph} l$ or  $\kappa \gg v_{\rm ph} l$ regimes providing subdominant contributions, with the growth rate roughly independent of $l$, and hence of $\kappa$.  

Another possibility is that the compressive component has a Kraichnan spectrum $W_{\rm 1D} (k) \propto k^{-3/2}$, as might be expected for compressive fast modes \citep{cho2003}. The case of {\it resonant} acceleration with a Kraichnan spectrum has been studied analytically, and it is a leading model for explaining giant radio halos in galaxy clusters \citep{Brunetti2007,miniati15}. However, our simulations in this paper have very limited dynamical range and do not resolve the fast mode cascade (which in any case is still very uncertain; see e.g. \citet{kowal10}, who find a Burgers-like fast mode spectrum). We therefore will compare our simulations to Equation \ref{eqn:growthPtuskin} appropriate for a Burgers-like spectrum.

Equations \ref{eqn1} - \ref{eqn:growthPtuskin} are derived assuming isotropic CR diffusion, but cross-field diffusion is much less efficient than field-aligned diffusion. Accounting for  anisotropic diffusion changes the above equations \citep{Chandran2004,Lynn2013} in the $\kappa >> v_{ph} L_{0}$ regime (Equation \ref{eqn1}), because CRs undergo a 1D random walk along magnetic field lines and are likely to return to the same eddy multiple times before the eddy is randomized in the turbulent flow. The characteristic timescale for turbulence-particle interactions is no longer the diffusion time across an eddy. Rather, it is the decorrelation time t$_{\rm corr} \sim L_0/v$ of the eddy, which is longer: $t_{\rm corr}/t_{\rm diff} \sim \kappa_{\parallel}/(v L_0) \gg 1$, when $\kappa_{\parallel} \gg v_{\rm ph} L_0$.  Over this time period, a typical CR can diffuse a distance $L_{B} \sim \sqrt{\kappa_{\parallel} t_{\rm corr}}$, over which time it will interact with $N \sim L_B/L_0 \sim \sqrt{\kappa_{||} t_{corr}}/L_0 \sim  \sqrt{\kappa_{||}/v L_{0}}$ eddies. Since over a period $t_{\rm corr} \sim (\kappa_{\parallel}/ v L_0) t_{\rm diff} \sim N^{2} t_{\rm diff}$ the CR spends time in $N$ eddies, on average it spends $t_{\rm eddy} \sim N t_{\rm diff}$ per eddy, i.e. it scatters with each eddy $N$ times, so that it acquires an rms momentum boost $(\delta p)_{\rm eddy} \sim N (\delta p)_{\rm iso}$.  
Thus, the momentum diffusion coefficient $D_{\rm pp} \sim (\delta p)_{\rm eddy}^{2}/t_{\rm eddy} \sim N^2 (\delta p)_{\rm iso}/(N t_{\rm diffuse}) \sim N D_{\rm pp,iso}
$, i.e. the reacceleration rate increases by a factor of $N$ compared to the momentum diffusion coefficient $D_{\rm pp,iso}$ that assumes isotropic spatial diffusion. Effectively, anisotropic diffusion increases the coherence of CR-turbulent scattering \citep{Chandran2004}: instead of momentum changes of order $(\delta p)_{\rm iso}$ on a timescale $t_{\rm diff}$, the CR makes larger momentum changes $(\delta p)_{\rm eddy} \sim N (\delta p)_{\rm iso}$ on longer timescales $t_{\rm eddy} \sim N t_{\rm diff}$. This is similar to reducing the opacity in radiative transfer, so that a photon has a longer mean free path and mean free time. Just as this reduces the photon escape time, the CR reacceleration time is reduced for a longer mean path in momentum space: $t_{\rm grow} \sim p^{2}/D_{\rm pp} \propto N^{-1}$, i.e.:
\begin{equation}
t_{\rm grow} \sim \frac{9}{2} \frac{\kappa_{\parallel}}{v^{2}} \left( \frac{v L_{0}}{\kappa_{\parallel}} \right)^{1/2}; \quad \kappa >> v_{ph}L_0
    \label{eqn:anisotropic}
\end{equation} 
This means that acceleration can still be reasonably efficient off the ``sweet spot" $\kappa_{\parallel} \sim v_{\rm ph} L_0$ for $\kappa_{\parallel} \gg v_{ph} L_0$, since $t_{\rm grow} \propto \kappa_{\parallel}^{1/2}$ instead of $t_{\rm grow} \propto \kappa$. On the other hand, there is little difference in acceleration rates between anisotropic or isotropic diffusion in the $\kappa << v_{ph} L_0$ regime.

To summarize: in the single eddy limit, assuming isotropic scattering, acceleration rates have a reasonably sharp peak at $\kappa \sim v_{\rm ph} L_0$, where $t_{\rm grow} \sim v_{\rm ph} L^{2}/v^{2}$ is minimized, and $t_{\rm grow} \propto \kappa $, $t_{\rm grow} \propto \kappa^{-1}$ in the $\kappa/v_{ph} L_0 \gg 1$, $\kappa/v_{ph} L_0 \ll 1$ regimes respectively. However, this dependence on $\kappa$ is modified by two effects. Firstly, a hierarchy of eddies contribute to reacceleration, and for $\kappa \ll v_{ph} L_0$, there exists a smaller eddy of scale $l$, such that $\kappa \sim v_l l$. For Burger's turbulence with sufficient dynamic range, $t_{\rm grow}$ becomes almost independent\footnote{It is important to appreciate that in our simulations, the limited dynamic range means that this effect does not really kick in -- acceleration rates are closer to the `single eddy' approximation, with $t_{\rm grow} \propto \kappa^{-1}$ in the $\kappa \ll v_{ph} L_0$ regime}, and hence we often still make use of the  `single eddy' formulae when comparing simulations with analytics. But in reality (or in simulations with much higher dynamic range), we expect a much milder dependence of acceleration rates on $\kappa$. of $\kappa$. Secondly, anisotropic diffusion increases the coherence of CR acceleration, so that $t_{\rm grow} \propto \kappa_{\parallel}^{1/2}$ increases more weakly with $\kappa_{\parallel}$ in the $\kappa \gg v_{ph} L_0$ regime. The net result is a characteristic minimum growth time $t_{\rm grow} \sim v_{\rm ph} L_0/v^{2}$ which depends only weakly on $\kappa$, in a range of several dex around the `sweet spot' $\kappa \sim v_{ph} L_0$.

\subsection{Non-Resonant Reacceleration of Streaming Cosmic Rays}
\label{sec:analytic_streaming}

While CRs in the $\sim$ GeV energy range are close to isotropic, even a small amount of drift anisotropy can excite magnetic perturbations through the resonant streaming instability \citep{Wentzel1968, Kulsrud1969, Skilling1971}. When the instability acts, forward traveling (relative to the CR drift) Alfv\'{e}n waves are most efficiently excited, with backward waves quickly damped. In the absence of wave damping, CRs pitch angle scatter off these waves until they isotropize in the wave frame, and thus advect along the magnetic field at the local Alfv\'{e}n speed. In the presence of wave damping, however, a steady-state balance between growth and damping gives a finite scattering rate which dictates that CRs diffuse relative to the wave frame. Since the streaming instability growth rate declines with increasing CR momenta, the diffusivity is also energy-dependent and generally rises with increasing CR energy. The resulting fluid CR transport is then a mixture of diffusion and streaming, as long as scattering by streaming generated waves dominates over scattering by extrinsic turbulence, a cut-off predicted to occur around 300 GeV \citep{Blasi2012, Kempski2021}, coincident with an observed change in the proton spectral index \citep{Evoli2019}. For energies above 300 GeV, CR propagation is thought to be purely diffusive.

While diffusion and streaming are similar in that they unlock CRs from the gas, the unique behavior of streaming fundamentally alters CR-wave interactions and, therefore, the reacceleration rate for CRs of any energy $E \lsim 300$ GeV. Let's consider a single compression. In the no transport case where CRs are perfectly locked to the gas, all energy gained via compression will be lost via rarefaction. Crucially, CR diffusion introduces a $\pi/2$ phase shift between CR pressure and gas density perturbations. This ``drag" against CRs provides a frictional force on compressive motions, giving rise to an acceleration that damps the wave. This phenomenon, akin to a damped simple harmonic oscillator, is known as Ptuskin damping \citep{ptuskin81} and is responsible for the non-resonant transfer of wave energy to CR energy (see \S \ref{sec:saturation} for additional discussion). 

The $\pi/2$ phase shift between CR and gas perturbations requires a CR flux $F_{\rm CR} \propto \nabla P_{\rm CR}$, such that the CR restoring force in response to perturbations is proportional to velocity (e.g., see \S2.1.1 of \citet{Tsung2021_staircase} for details), which is the requirement for either damping or driving. On the other hand, the CR flux with pure streaming is $F_{\rm CR} \propto P_{\rm CR}$, resulting in a CR restoring force proportional to displacement and therefore unable to create a $\pi/2$ phase shift necessary for Ptuskin damping and reacceleration -- as in a simple harmonic oscillator, energy is conserved. With pure streaming and no diffusion, there is then no energy transfer from waves to CRs.

Another way to understand this is in terms of phase space transport. Spatial diffusion is tied to momentum diffusion (as can be seen from the relationship between the spatial ($\kappa$) and momentum ($D_{\rm pp}$) diffusion coefficients, e.g. in equation \ref{eq:Dpp_Wk}), while spatial advection is tied to momentum advection (as is true, for instance, in adiabatic compression or expansion). Since there is more phase space at higher momenta, CRs diffusing in phase space inevitably diffuse to higher energy. By contrast, in CR advection, there is no stochasticity: CR evolution is adiabatic in the Alfv\'{e}n wave frame. This can be seen in the equation for the distribution function $f(x,p)$ \citep{Skilling1971}:
\begin{equation} 
\frac{Df}{Dt} \equiv \left( \frac{\partial}{\partial t} + \mathbf{w} \cdot \nabla \right) \, f = \frac{1}{3} ( \nabla \cdot \mathbf{w} ) p \frac{\partial f}{\partial p}
\end{equation} 
where $\mathbf{w} = \mathbf{v} + \mathbf{v_{\rm A}}$ is the net velocity of Alfv\'{e}n waves in the lab frame. Since $\langle \nabla \cdot \mathbf{w} \rangle =0$,  i.e. there is no net converging or diverging flow for Alfv\'{e}n waves over a box which is homogeneous on large scales, $\langle Df/Dt \rangle =0$ -- and a time-stationary distribution function means no acceleration is taking place. All energy changes are reversible, since the energy gained due to converging Alfv\'{e}n waves during compression is returned to the plasma as CRs stream out of the overdensity, resulting in diverging Alfv\'{e}n waves. Advection only produces a net energy change when $\langle \nabla \cdot \mathbf{w} \rangle \neq 0$ -- for instance, when there is net compression or expansion of the fluid $\langle \nabla \cdot \mathbf{v} \rangle \neq 0$, or there is a net change in Alfven speed, due to net gradients in B-field strength or gas density $\langle \nabla  \cdot \mathbf{v_{\rm A}} \rangle \neq 0$. The latter happens, for instance, when CRs stream in a stratified medium, resulting in CR energy loss at a rate $\mathbf{v_{\rm A}} \cdot \nabla P_c$.

So what is the role of streaming in CR-wave interactions? Streaming introduces two additional effects. First, the advective transport modifies the perturbed acceleration from the phase-shifted CR force. One can derive the acceleration rate $\dot{u}$ in the diffusion only and then streaming-modified cases \citep{Begelman1994, Tsung2021_staircase}:

\begin{equation}
    \frac{\dot{u}}{u} \sim -\frac{c_{c}^{2}}{\kappa}  \quad \rm (diffusion) 
    \label{eqn:phaseshift_diff}
\end{equation}
\begin{equation}    
    \frac{\dot{u}}{u} \sim -\frac{c_{c}^{2}}{\kappa} \left(1 \mp \frac{v_{A}}{c_{s}} \right) \quad \rm (diffusion + streaming)
    \label{eqn:phaseshift_stream}
\end{equation}

where $u$ is the velocity, $c_{\rm c} \sim (P_{\rm CR}/\rho)^{1/2}$, and $\Gamma \sim \dot{u}/u$ is the damping/growth rate of gas motions. Note that only the streaming-dominated case allows for growth ($\Gamma > 0$). If there were a background CR gradient much larger than the perturbed gradients due to gas motions, growth can occur in the diffusion-only case when the diffusion time is shorter than the sound crossing time across a CR scale length $L_{c}$ \citep{DruryFalle1986}; however, that is not the case for the unstratified medium we assume here. In the streaming case, though, unstable growth is possible, regardless of $L_{c}$, if $\beta \ll 1$. This case was studied in detail in \citet{Tsung2021_staircase}. For now, we will operate in the $\beta \ge 1$ regime. Note also that Equation \ref{eqn:phaseshift_stream}, as written, assumes an isothermal equation of state where gas heating by CRs is neglected. In non-isothermal gas, Equation \ref{eqn:phaseshift_stream} is multiplied by $1 \pm (\gamma_g - 1)v_{A}/c_{s}$ \citep{Begelman1994}.

The $\pm$ in Equation \ref{eqn:phaseshift_stream} refers to gas motion in same (+) or opposite (-) direction relative to the direction of cosmic ray drift. Overdensities are created by converging gas flows, but CRs stream out of them in the opposite direction. These opposing flows reduce the net amount of CR compression in the Alfven wave frame, and hence reduce CR energization. Thus, the `-' sign is appropriate, and CRs take energy from gas motions at a new, reduced rate modified by the factor $1 - \frac{v_{A}}{c_{s}}$. The new CR reacceleration time relative to the pure diffusion growth time, obtained by dividing Equation \ref{eqn:phaseshift_diff} by Equation \ref{eqn:phaseshift_stream}, is then

\begin{equation}
    \frac{t_{\rm grow}^{\rm stream}}{t_{\rm grow}^{\rm diff}} \sim \frac{1}{1 - \sqrt{2/ \beta}} \quad (\kappa < v_{ph} L_{0})
    \label{eqn:streamGrowthMod}
\end{equation}

Note that the growth rate now depends on plasma $\beta$, which for an isothermal ($\gamma_{g} = 1$) gas is $\beta = 2c_{s}^{2}/v_{A}^{2}$. For an adiabatic $\gamma_g = 5/3$ gas, the growth rate is increased by a factor of $(1 + 2v_{A}/3c_{s})$, still resulting in a significant decrease in reacceleration when $v_{A} \sim c_{s}$. As $\beta$ increases, the rate converges to the pure diffusion rate. Note also that Equation \ref{eqn:streamGrowthMod} is only appropriate in the regime $\kappa < v_{ph} L_{0}$. For $\kappa >> v_{ph} L_{0}$, $t_{\rm grow} \propto \kappa_{||}^{1/2}$ instead of $t_{\rm grow}  \propto \kappa_{||}$ (Equation \ref{eqn:anisotropic}). With our streaming modification included, we can write $\kappa_{\rm eff} \rightarrow \kappa/(1 - \sqrt{2/ \beta})$, and the new $t_{\rm grow}^{\rm stream}$ should be $\propto 1/(1 - \sqrt{2/ \beta})^{1/2}$ instead of $\propto 1/(1 - \sqrt{2/ \beta})$.

\begin{equation}
    \frac{t_{\rm grow}^{\rm stream}}{t_{\rm grow}^{\rm diff}} \sim \left(\frac{1}{1 - \sqrt{2/ \beta}}\right)^{1/2} \quad (\kappa >> v_{ph} L_{0})
    \label{eqn:streamGrowthMod_largeDiff}
\end{equation}

Second, wave excitation by the streaming instability drains energy from the CR population at a rate H = $v_{A} \cdot \nabla P_{\rm CR}$ \citep{ZweibelReview2017}. This energy is subsequently transferred to the gas by wave damping. For an isothermal equation of state, the gas does not gain energy, but the energy sink for the CRs remains and competes with compressional heating to decrease the CR reacceleration rate. As we'll see in \S \ref{sec:heating}, however, reacceleration rates are not predominantly stunted by streaming energy loss; instead, when reacceleration rates are slowest (the low-$\beta$ regime), CR energy loss rates are also slowest. The major correction to growth rates appears to come from the modified phase-shifted CR force.

\section{MHD Simulations}
\label{sec:numerics}

Motivated by the above considerations, the agenda for numerical simulations is straightforward:
\begin{itemize} 
\item{Study growth times in the pure diffusion case as a function of $\kappa$, and compare to Ptuskin's predictions (Equations \ref{eqn1}-\ref{eqn:tmin_grow}, \ref{eqn:growthPtuskin}).} 
\item{Study the effect of anisotropic diffusion, which is expected to reduce $t_{\rm grow} \rightarrow t_{\rm grow}/(\sqrt{\kappa/v L_0})$ in the $\kappa \gg v_{\rm ph} L_0$ regime.}
\item{Study the combined effect of CR streaming and diffusion on turbulent reacceleration, and check our new analytic expectations (Equations \ref{eqn:streamGrowthMod}, \ref{eqn:streamGrowthMod_largeDiff}), particularly as a function of plasma $\beta$. When streaming is important, we expect acceleration to be inefficient at low $\beta$.} 
\item{Quantify the fractional turbulent dissipation going into CRs, and the non-linear saturation of CR reacceleration.}

\item{Study the resolution dependence of both CR reacceleration and streaming energy loss. Are these well captured in zoom-in cosmological or even fully cosmological simulations of galaxy formation?  
We show additional convergence tests and discuss implications in  Appendix \ref{appendix:convergence}.}  
\end{itemize}

\subsection{Computational Methods}

We ran a suite of simulations using a version of the magnetohydrodynamics (MHD) code Athena++ \citep{AthenaRef} modified to include CRs \citep{JiangCRModule}. CR transport via diffusion and streaming is implemented via a two-moment method developed originally for radiative transfer, and the efficiency and accuracy of this implementation have been extensively tested \citep{JiangCRModule}.

The equations solved are a combination of the ideal MHD equations and CR evolution equations for a mixture of streaming and diffusive transport: 

\begin{equation*}
    \frac{\partial \rho }{\partial t} + \nabla \cdot (\rho \mathbf{v_{g}}) = 0 \quad \quad \frac{\partial \mathbf{B} }{\partial t} - \nabla \times (\mathbf{v_{g}} \times \mathbf{B}) = 0 
\end{equation*}

\begin{equation*}
\begin{split}
    \frac{\partial \rho \mathbf{v_{g}} }{\partial t} + \nabla \cdot (\rho \mathbf{v_{g}} \mathbf{v_{g}} - \mathbf{B} \mathbf{B} + \mathbf{I}\left(P_{\rm g} + P_{B}\right) = \\
    \sigma_{c} \cdot [\mathbf{F_{\rm CR}} - \mathbf{v_{g}} \cdot (E_{\rm CR}\mathbf{I} + \mathbf{P_{\rm CR}})]
\end{split}
\end{equation*}

\begin{equation*}
\begin{split}
\frac{\partial E }{\partial t}  + \nabla \cdot \left[(E + P_{\rm g} + P_{B}) \mathbf{v_{g}} - \mathbf{B}(\mathbf{B} \cdot \mathbf{v_{g}}) \right] = \\ (\mathbf{v_{g}} + \mathbf{v_{st}}) \cdot (\mathbf{\sigma_{c}} \cdot [\mathbf{F_{\rm CR}} - \mathbf{v_{g}} \cdot (E_{\rm CR}\mathbf{I} + \mathbf{P_{\rm CR}})])
\end{split}
\end{equation*}

\begin{equation*}
\begin{split}
\frac{\partial E_{\rm CR} }{\partial t} + \nabla \cdot \mathbf{F_{\rm CR}} = - (\mathbf{v_{g}} + \mathbf{v_{st}})  \cdot \\ (\mathbf{\sigma_{c}} \cdot [\mathbf{F_{\rm CR}} - \mathbf{v_{g}} \cdot (E_{\rm CR}\mathbf{I} + \mathbf{P_{\rm CR}})])
\end{split}
\end{equation*}

\begin{equation*}
\frac{1}{v_{m}^{2}} \frac{\partial F_{\rm CR} }{\partial t} + \nabla \cdot \mathbf{P_{\rm CR}} = -\mathbf{\sigma_{c}} \cdot [\mathbf{F_{\rm CR}} - \mathbf{v_{g}} \cdot (E_{\rm CR}\mathbf{I} + \mathbf{P_{\rm CR}})]
\end{equation*}

where $\rho$ is the gas density, $\mathbf{v_{g}}$ is the gas velocity, $\mathbf{v_{st}} = - \mathbf{v_{\rm A}} (\mathbf{B} \cdot \nabla P_{\rm CR})/|\mathbf{B} \cdot \nabla P_{\rm CR}|$ is the CR streaming velocity, $\mathbf{B}$ is the magnetic field, $E$ is the thermal energy, $E_{\rm CR}$ is the CR energy, $F_{\rm CR}$ is the CR energy flux, and, $P_{B} = B^{2}/8\pi$, $P_{\rm CR} = (\gamma_{\rm CR} - 1) E_{\rm CR}$ are the magnetic and CR pressures. 

Note that the CR adiabatic index is $\gamma_{\rm CR} = 4/3$, and we use an isothermal equation of state, for which the gas adiabatic index is $\gamma_{g} = 1$. While gas can be \emph{effectively} isothermal when radiative cooling is strong, this isothermality assumption is not rigorously appropriate. However, since the reacceleration time for a fixed $\kappa$ depends on the phase velocity of compressible waves, i.e. the sound speed, we enforce $\gamma_g = 1$ to keep a constant phase velocity and facilitate an easier comparison to analytic expectations. Our conclusions and applicability to the ISM, CGM, and ICM are quite insensitive to this choice: note that the minimum growth time from Equation \ref{eqn:growthPtuskin} scales as $v_{ph} \propto \sqrt{\gamma_{g}}$ and, hence, changes by the small factor of $\sqrt{5/3}$ modulating between an isothermal and adiabatic equation of state. An alternative implementation, with $\gamma_{g} = 5/3$ and including radiative cooling, requires an additional heating term tuned to enforce global thermal balance (rather than allowing for a cooling runaway), but this heating term is often invoked as a substitute or rough parametrization of CR heating itself. As a natural starting place for our study of cosmic ray - turbulence interplay, the isothermality assumption is cleaner. 

The two-moment method presents itself through the inclusion of a maximum speed of light parameter $v_{m}$ and an interaction coefficient (Equation 10 of \cite{JiangCRModule}) $\mathbf{\sigma_{c}}^{-1} = \mathbf{\kappa} +  \mathbf{v_{st}} \cdot (E_{\rm CR}{\bf I} + {\bf P_{\rm CR}})$. $\mathbf{\kappa}$ is the CR diffusivity. Source terms in the momentum and energy equations depend on this interaction coefficient and encapsulate how CRs exchange momentum and energy with the gas. In the gas thermal energy and CR energy equations, these source/sink terms account for collisionless energy transfer from the CRs to the thermal gas due to wave damping. We will sometimes refer to this collisionless energy transfer as ``streaming energy loss." Time-dependent hydromagnetic wave energy is not explicitly tracked here, as streaming instability growth times are generally much shorter than other timescales of interest; waves are assumed to couple CRs to gas unless $\nabla P_{\rm CR} \rightarrow 0$ (see \citealt{Thomas2019} for an implementation which tracks wave energy).

\subsection{Generating Turbulence} 
To generate turbulence, we take advantage of the turbulent stirring module in Athena++, which uses an Ornstein-Uhlenbeck process \citep{Uhlenbeck1930} to smoothly generate a prescribed mixture of compressive and solenoidal velocity perturbations $\hat{v} = f_{\rm shear} \hat{v}_{\rm shear} + (1-f_{\rm shear}) \hat{v}_{\rm compressive}$ over a correlation time (similar methods are employed in e.g. \citealt{EswaranPope1988, Federrath2008, Federrath2010, Lynn2012}). The turbulent reacceleration rate of CRs depends only on the compressive component of turbulence. For this paper, we will focus on purely compressive forcing ($f_{\rm shear} = 0$); increasingly solenoidal forcing leads to weaker compressions and rarefactions for a given Mach number, therefore decreasing the turbulent reacceleration rate further until it becomes zero with purely solenoidal perturbations. The advantage of purely compressive driving is that the turbulent dynamo (which depends on solenoidal driving) does not operate, so our simulations have roughly constant plasma $\beta$. Otherwise, since we need to drive the simulations for many eddy turnover times to see CR turbulent reacceleration, turbulent magnetic field amplification would obscure the plasma $\beta$ dependence of CR reacceleration that we wish to study. Instead of enforcing a specific turbulent power-law over many scales, we use parabolic driving between modes $1 < k < 3$, and the resulting turbulent cascade to higher wavenumbers is created organically. For driving, we set the autocorrelation timescale to be $t_{\rm corr} =  L/c_{s}$ and drive fluctuations every $t_{\rm drive} = 2 \times 10^{-3} (L/c_{s})$. Our results are not sensitive to these assumptions. 

\begin{table}
  \centering
  \caption{Range of simulation parameters used in \S \ref{sec:purediff}, \ref{sec:stream}. In \S \ref{sec:purediff}, to check analytic growth predictions, we fix $\beta = 2$, the initial $P_{\rm CR}/P_{\rm g} = 100$, and the grid size to $64^{3}$. We test convergence with respect to grid size and maximum speed of light, $v_{m}$, in the Appendix. In \S \ref{sec:stream}, all parameters are the same, but we vary $\beta$ between 2 and 200, and we run a set of higher resolution simulations with grid sizes of $128^{3}$ -  $512^{3}$.}
  \begin{tabular}{l|r}
    \toprule
    Box Size & $(2L)^{3}$ \\
    Resolution & $2L/64$ - $2L/512$  \\
    Outer driving scale, $L_0$ & $\sim 2L/3$ \\
    $\beta = P_{\rm g}/P_{B}$ & 2.0 - 200 \\
    $P_{\rm g}/P_{\rm CR}$ & 100 \\
    $t_{\rm drive}$ & $2 \times 10^{-3} (L/c_{s})$ \\
    $t_{\rm corr}$ & $L/c_{s}$ \\
    $\mathcal{M}_{\rm s,turb}$ & 0.25, 0.35, 0.5 \\
    $v_{m}/c_{s}$ & 50 - 400 \\
  \end{tabular}
\label{table1}
\end{table}

To check simulated growth times vs analytic predictions, we use grids with $64^{3}$ - $256^{3}$ cells in a square domain of length 2L, and we measure the outer-scale turbulent eddy (the one with the most power) to be of size $\sim 2L/3$. This low resolution gives us only a short turbulence inertial range -- dissipation sets in at a scale $\approx 30$ cell widths \citep{Federrath2010, Federrath2020} -- but in practice, we find that even our lowest resolution runs with a $64^{3}$ box give reasonably converged reacceleration rates matching analytic expectations\footnote{Note that these CR hydrodynamics simulations, with our fiducial choice of $v_{m}$, are about 8 times more expensive than MHD simulations, which are $\sim $2x more expensive than pure hydro simulations. This is due partially to additional overhead from the CR module but primarily due to the maximum speed of light $v_{m}$, which should be much faster than other MHD propagation speeds in the system and sets the timestep. Our $256^{3}$ and $512^{3}$ simulations, then, are about as expensive as $512^{3}$ and $1024^{3}$ hydro runs, typical for parameter scans and production runs in the turbulence literature.}. This allows us to run a large parameter study to verify the scalings of Section \ref{sec:background}.

In \S\ref{sec:stream}, we find that we must be a bit more careful with our resolution and simulation box sizes. While reacceleration rates are again converged even at our lowest resolution of $2L/64$, streaming energy loss rates in low-$\beta$ plasmas are sensitive to the amount of magnetic field tangling captured in the simulation; to show this, we vary some of our simulations to have resolutions of $2L/128$, $2L/256$, and $2L/512$. For each simulation, we choose a fiducial maximum speed of light $v_{m} = 50 c_{s}$, and we show convergence with respect to this choice in the Appendix. Table \ref{table1} compiles our fiducial parameters. In Section \ref{sec:stream}, all parameters are the same, except we vary the initial $\beta$ between 2 and 200. 

\begin{figure*}
\centering
\includegraphics[width=0.75\textwidth]{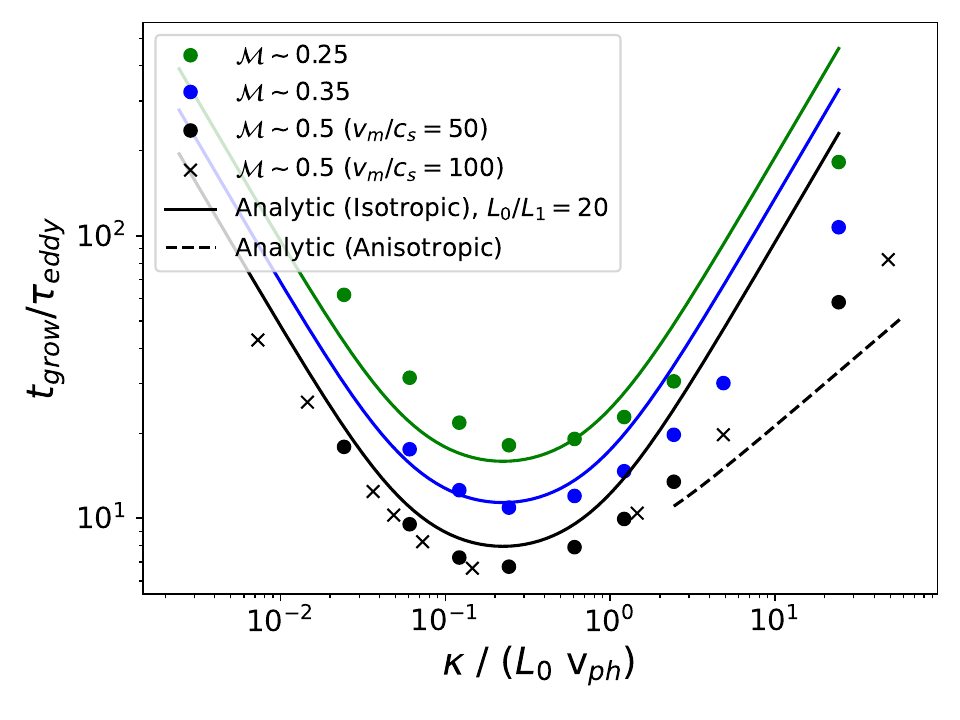}
\caption{CR energy growth times, normalized by the outer scale eddy turnover time ($\tau_{\rm eddy}$) for simulations each with $\beta = 1$ but varying $\mathcal{M}$ (different colors). Simulations were run using either $v_{m}/c_{s} = 50$ (dots) or $v_{m}/c_{s} = 100$ (x's); see the Appendix for a full comparison of different $v_{m}$ and different resolutions. Lines show analytic expectations assuming isotropic diffusion (solid  lines, \citealt{Ptuskin1988}) and including a correction for anisotropic diffusion (dashed line, \citealt{Chandran2004}) for $\kappa > v_{ph} L_{0}$. Simulations match expectations remarkably well for $\kappa < v_{ph} L_{0}$ if one sets $L_{0}/L_{1} = 20$, i.e. if $L_{1}$ represents one cell width. Simulations also reflect the expected decrease in growth time due to anisotropic diffusion.}
\label{fig:GrowthTimes_Ptuskin1988}
\end{figure*}

\subsection{Results: Pure Diffusion}
\label{sec:purediff}

We begin with simulations without streaming, i.e. solely anisotropic diffusion along the local magnetic field direction. Each simulation has $\beta = 2$ and low initial CR pressure ($P_{\rm g}/P_{\rm CR} = 100$) so the phase velocity is approximately the isothermal sound speed, $c_{s}$. We choose three different forcing amplitudes to give turbulent Mach numbers of $\mathcal{M} \sim 0.25$, $0.35$, and 0.5 (measured at late times when the RMS velocity has saturated), and we vary the diffusion coefficient over more than 3 orders of magnitude to test whether we recover the growth time prediction of Equation \ref{eqn:growthPtuskin}.

The results are plotted in Figure \ref{fig:GrowthTimes_Ptuskin1988}. The simulation growth times are calculated by summing the total CR energy in the box, $E_{\rm CR}^{tot}$, and fitting a line to the ${\rm log}_{10}(E_{\rm CR}^{tot})$ curve in the exponential growth phase between 3 and 6 eddy turnover times. This time interval was chosen because it spans many time outputs and occurs after the kinetic energy has saturated but before the CR energy rises past equipartition with the thermal energy, at which point changes to the gas compressibility (changes to v$_{ph}$) decrease the growth rate and affect the normalization of $\kappa$ relative to $L_{0} v_{ph}$. A rolling derivative confirms that these growth times are representative of the exponential growth phase. 

Our simulations match analytic expectations (Equation \ref{eqn:growthPtuskin}) very well, not only reproducing the correct scaling with Mach number ($t_{\rm grow}/\tau_{\rm eddy} \propto 1/\mathcal{M}$) but also the correct growth curve shape in the $\kappa \leq v_{ph} L_0$ regime if one sets $L_0/L_1 \approx 20$. The appropriate value of $L_0/L_1$ is determined, in reality, by the characteristic width of a shock front, $L_1$, relative to the outer driving scale, $L_0$. In simulations, however, $L_1$ is limited by resolution; our choice of $L_0/L_1 \approx 20$ is motivated by $L_1$ spanning roughly one cell width in our $64^{3}$ grid. In the Appendix, we show how other choices of $L_0/L_1$, which should correspond to other spatial resolutions, change the growth curves.

In the $\kappa > v_{ph} L_0$ limit, we also recover the expected decrease in growth time due to anisotropic rather than isotropic diffusion \citep{Chandran2004}: $t_{\rm grow} \rightarrow t_{\rm grow}/(\sqrt{\kappa/v L_0})$. The dashed line in Figure \ref{fig:GrowthTimes_Ptuskin1988} shows $t_{\rm grow}/\tau_{\rm eddy}$ in the $\kappa >> v_{ph} L_0$ regime when correcting for anisotropic diffusion, and our simulated growth times match this trend very well.

\subsection{Results: With Streaming}
\label{sec:stream}

We now include streaming transport and a variety of initial plasma $\beta \in [2, 200]$ to test Equation \ref{eqn:streamGrowthMod} in the streaming-dominated regime, as well as the predicted collapse of $t_{\rm grow}^{\rm stream} \rightarrow t_{\rm grow}^{\rm diff}$ in the $\kappa >> c_{s} L_{0}$ regime. We vary the diffusion coefficient between $\kappa = 0.15 L_0 c_{s}$ (the maximal growth case without streaming) and $\kappa = 15 L_0 c_{s}$.

An important note is that we use the same forcing for each simulation, i.e. $\epsilon = \rho v^{3}/L_0$ is held constant; therefore, for increasing plasma $\beta$, the magnetic field back-reacts on the flow less, leading to a slightly higher turbulent Mach number and average velocity divergence. This is a mild effect. Nonetheless, to make consistent comparisons, we run each simulation with and without streaming for each plasma $\beta$ and focus our analysis on the \emph{ratio} of growth rates. As in Section \ref{sec:purediff}, the initial CR pressure is $1\%$ of the thermal pressure, so CRs don't significantly affect the properties of turbulence. 

\begin{figure*}
\centering
\includegraphics[width=0.48\textwidth]{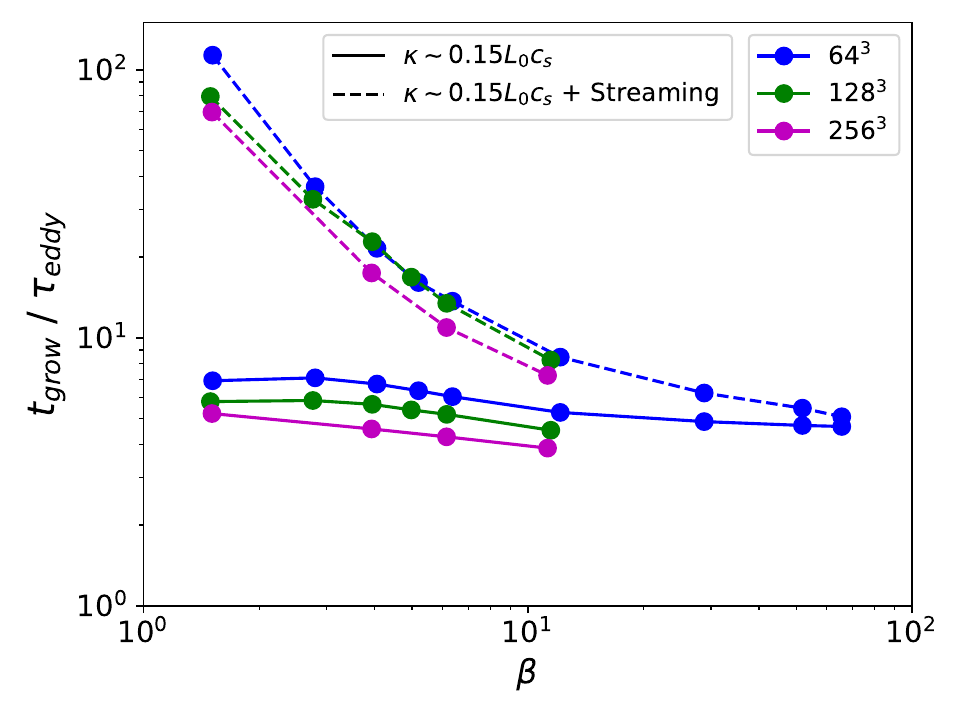}
\includegraphics[width=0.48\textwidth]{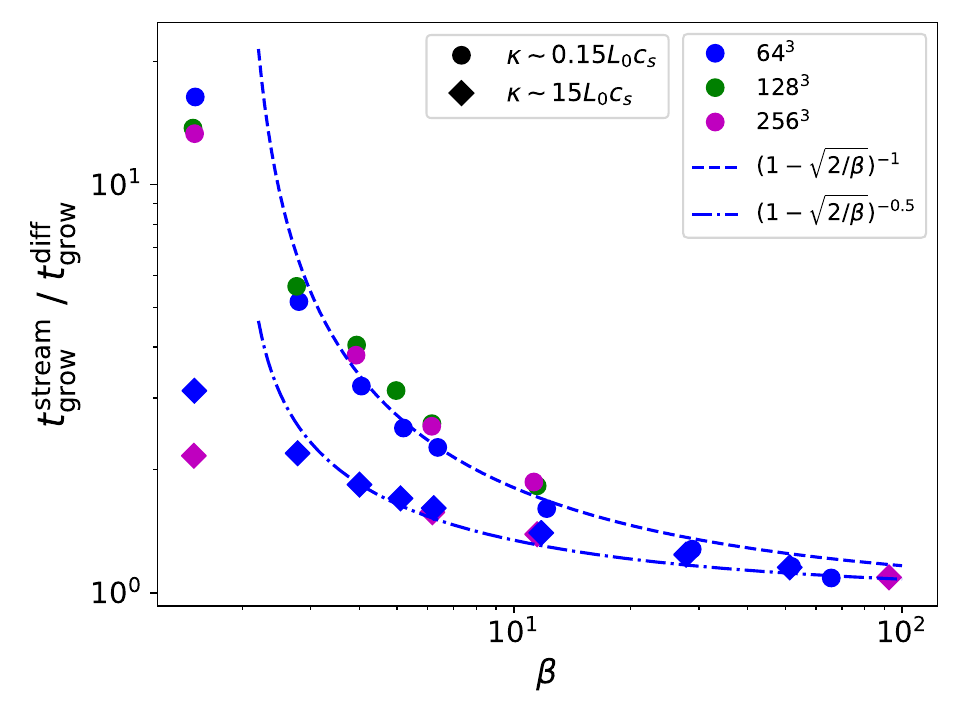}
\caption{\emph{Left:} Growth times as a function of plasma $\beta$ for CRs with (dashed lines) and without (solid lines) streaming, both with parallel diffusion coefficients of $\kappa_{||} = 0.15 c_{s} L_{0}$ such that the growth time is near the minimum. Different colors correspond to different grid sizes. With only diffusion, growth times are near the minimum value (see Figure \ref{fig:GrowthTimes_Ptuskin1988}), but streaming losses significantly offset the energy gain when $\beta$ is low. \emph{Right:} Ratio of growth times with and without streaming. The dashed blue line shows the analytic estimate for the $\kappa < v_{ph} L_{0}$ regime (Equation \ref{eqn:streamGrowthMod}), and the blue dot-dashed line shows the analytic estimate for the $\kappa >> v_{ph} L_{0}$ regime (Equation \ref{eqn:streamGrowthMod_largeDiff}). Compared to canonical rates assuming pure diffusion, reacceleration is less efficient in low-$\beta$ environments when streaming is included. }
\label{fig:CR_Energy_betaStudy}
\end{figure*}

Our results are shown in Figure \ref{fig:CR_Energy_betaStudy}. The left panel shows growth times for streaming and non-streaming simulations. These are normalized by the outer scale eddy turnover time, which assumes $\mathcal{M} \sim 0.5$ for each run, but decreased growth times for pure diffusion at higher $\beta$ reflect the reduced MHD forces on the flow, leading to $\mathcal{M} > 0.5$. The x-axis denotes the \emph{evolved} plasma $\beta$ of the simulation; because our forcing has no solenoidal component, magnetic field amplification is relatively inefficient, and plasma $\beta$ decreases by a factor less than 2 during the time interval of our analysis. 

The right panel shows the ratio of reacceleration times for anisotropically diffusing CRs, with and without additional streaming. At low $\beta$, typical of the ISM, the growth time is an order of magnitude longer than the growth time with pure diffusion. The discrepancy decreases at higher $\beta$ but is still a factor of $\sim 2$ even at $\beta \sim 10$. The dashed blue line shows the expected ratio in the $\kappa < v_{ph} L_{0}$ regime (Equation \ref{eqn:streamGrowthMod}). Indeed, our results broadly fit with expectations. When streaming is important ($\kappa = 0.15 L_{0} c_{s}$), energy gains are largely reversible, and growth times are much longer at low $\beta$. At higher $\beta$, the discrepancy drops, following the predicted scaling of Equation \ref{eqn:streamGrowthMod} fairly well. For $\kappa = 15 L_{0} c_{s}$, the effects of streaming are lower, following the expectation from Equation \ref{eqn:streamGrowthMod_largeDiff} for the $\kappa >> v_{ph} L_{0}$ regime (the blue dot-dashed line). Note that growth times in this regime, even without additional streaming modifications, are already much longer than the minimum growth time when $\kappa < L_{0} c_{s}$. Overall, the main point is clear: \emph{CR streaming alters CR-turbulence interactions and significantly decreases reacceleration rates in low-$\beta$, ISM-like environments compared to the pure diffusion growth rates first derived in \cite{Ptuskin1988} and frequently assumed in CR propagation models.}

\subsection{Saturation}
\label{sec:saturation}

\begin{figure}
\centering
\includegraphics[width=0.45\textwidth]{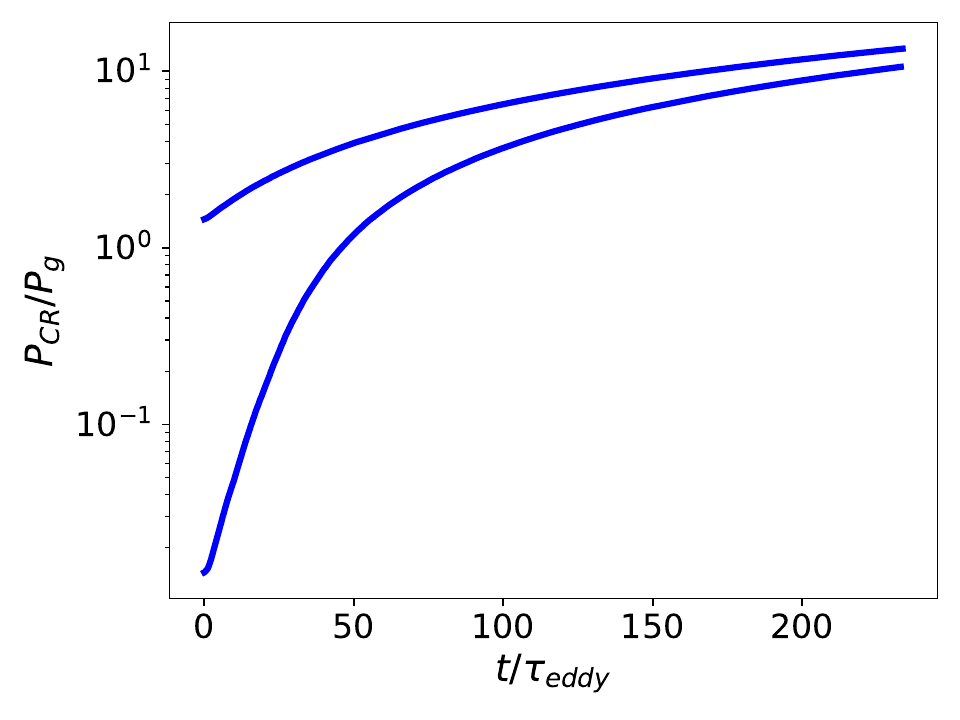}
\includegraphics[width=0.45\textwidth]{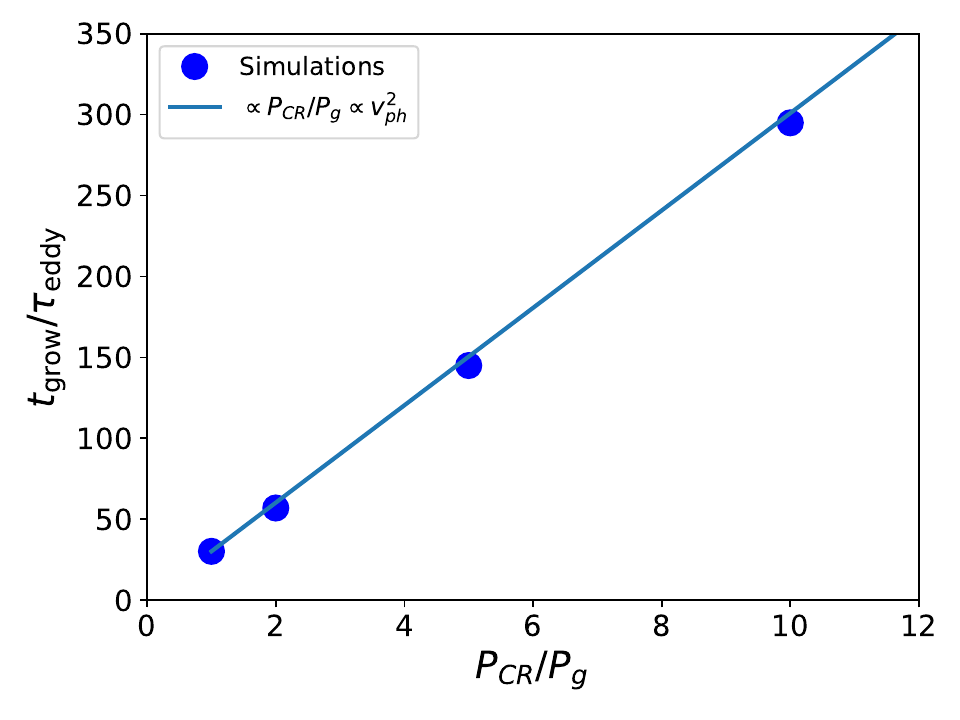}
\caption{\emph{Top:} Two representative simulations starting from $P_{CR}/P_{g}$ = 1/100 and 1, run out to hundreds of eddy turnover times. 
Growth is exponential, with increasing growth times and a transition to quasi-linear growth once $P_{\rm CR} > P_{\rm g}$. 
\emph{Bottom:} Simulated growth time vs $P_{\rm CR}/P{g}$ for simulations with constant $\kappa < v_{ph} L_{0}$ and varying $P_{\rm CR} > P_{g}$. We expect (and find) that $t_{\rm grow} \propto v_{ph}^{2} \propto P_{\rm CR}/P_{g}$, consistent with Equation \ref{eqn2}.}
\label{fig:CR_Saturation}
\end{figure}

Now that we've analyzed the linear regime when $P_{CR} << P_{g}$ and compared to analytic growth rates, we move to the nonlinear regime when CRs become dynamically significant. In principle, CRs can back-react on the turbulent flow, changing its energy spectrum and cascade rates, much as for instance magnetic fields alter hydrodynamic turbulence, resulting in MHD turbulence. A fuller exploration of this interesting issue requires higher resolution simulations and is the subject of our follow-up paper (Bustard and Oh, in prep). Given our focus on energetics, however, we can at least study in this work how turbulent kinetic energy is dissipated. In hydrodynamics, the turbulent energy forcing rate $\tilde{\epsilon} \approx \rho v^{3}/l \approx$const (independent of scale $l$) is equal to the gas heating rate. In MHD, some fraction of the kinetic energy is used to amplify magnetic fields via the turbulent dynamo. Similarly, in a two-fluid CR-gas system, some fraction of the kinetic energy does not dissipate into heat at small scales, but is instead diverted through the CR population. 

Since $\dot{P}_{\rm CR} \propto P_{\rm CR}/t_{\rm grow}$, where $t_{\rm grow}$ is initially independent of $P_{\rm CR}$, we expect exponential growth in $P_{\rm CR}(t)$, similar to the magnetic turbulent dynamo where $\dot{P}_{\rm B} \propto \omega P_{\rm B}$ (where $\omega$ is the fluid vorticity), and magnetic fields initially grow exponentially. For the turbulent dynamo, the fluid vorticity decreases due to magnetic tension from the growing magnetic field, and the dynamo transitions from exponential to linear growth, and finally saturation at roughly equipartition values. 

The non-linear saturation of CR turbulent reacceleration is also interesting. Presumably, for fixed driving, as $P_{\rm tot}$ increases, the fluid becomes less compressible due to an increase in $v_{\rm ph}$, which decreases $M_{\rm ph}$ and increases $t_{\rm grow}$. While saturation in the turbulent dynamo is due to a decrease in $\omega = \nabla \times v$, it is related to a decrease in $\nabla \cdot v$ for CR reacceleration. This holds true in our simulations. In Figure \ref{fig:CR_Saturation}, we compare simulated growth times for diffusion-only simulations with $\kappa = 0.15 L_{0} c_{s}$ and varying $P_{\rm CR}/P_{g} > 1$. We find, as expected given Equation \ref{eqn2}, that due to the rise in $v_{\rm ph}$ due to the increase in CR energy density, the growth time increases secularly: $t_{\rm grow} \propto v_{ph}^{2} \propto P_{\rm CR}/P_{g}$. Thus, $\dot{P_{\rm CR}} \propto P_{\rm CR}/t_{\rm grow} \rightarrow$ constant, and growth transitions from exponential to linear, just as for the magnetic turbulent dynamo. Note that this saturation only occurs in the $\kappa < v_{\rm ph} L$ regime and once $P_{\rm CR} \ge P_{g}$. Otherwise, there is little change in $t_{\rm grow}$.

\section{Suppression of Streaming Energy Loss}
\label{sec:heating}
\begin{figure*}
\centering
\includegraphics[width=0.8\textwidth]{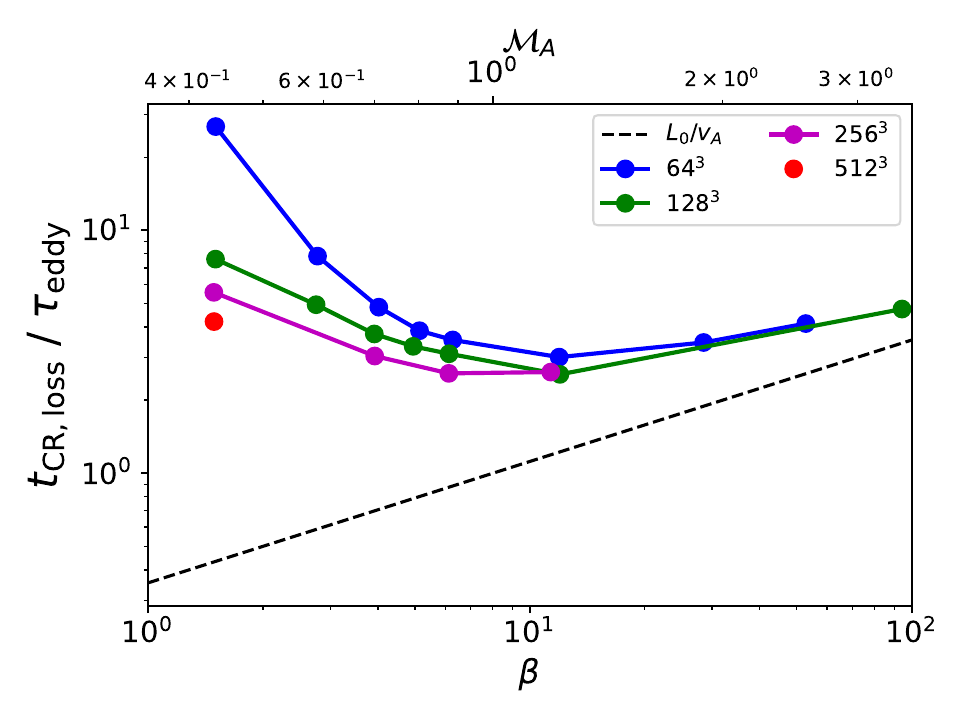}
\caption{\emph{Top:} CR streaming energy loss time (in units of the eddy turnover time) compared to the approximation $L_0/v_{A}$ (dashed line). Each simulation has $M_{\rm s} \approx 0.5$ and streaming-dominated transport, with $\kappa \sim 0.15 L_{0} c_{s}$, hence $L_{0}/v_{A} = \tau_{\rm stream} < L_{0}^{2}/\kappa = \tau_{\rm diff}$ for $\beta \lessapprox 100$. At high-$\beta$, turbulence easily tangles field lines, and the energy loss time approaches $L_{0}/v_{A}$. At low-$\beta$, when there is a strong guide field, field line tangling is more difficult, and misalignment between the magnetic field and the CR pressure gradient suppresses energy loss by a factor $\sim 0.1$. The amount of energy loss is resolution-dependent for low $\beta$.}
\label{fig:heat_suppression}
\end{figure*}

We additionally quantify how magnetic field strength affects the streaming energy loss rate. This can be thought of interchangeably as the gas heating rate due to CR streaming, if the gas equation of state is adiabatic, but here no heating occurs because the gas is isothermal. Figure \ref{fig:heat_suppression} shows the collisionless energy loss time, calculated as $t_{\rm CR, loss} = P_{\rm CR} / |v_{A} \cdot \nabla P_{\rm CR}|$ for simulations with varying plasma $\beta$. It also shows the naive expectation that, if the CR scale height is approximately the outer eddy scale $L_{0}$, the relative loss time should be $L_{0}/v_{A}$, so that loss times decrease monotonically for stronger B-fields. In fact, that is far from the case. At low $\beta$, the loss time is orders of magnitude longer, rising with increasing resolution at low $\beta$ but seemingly converged with resolution for $\beta > 10$. The loss time reaches a minimum near $\beta \sim 10$, which corresponds to an Alfv\'{e}n Mach number $M_{A} = v/v_{A} \sim 1$. Note also that streaming energy loss time is \emph{not} inversely correlated with the reacceleration time, as it would be if streaming energy loss were the main factor stunting reacceleration. Instead, slow reacceleration in the low-$\beta$ regime is also accompanied by slow energy loss, pointing to the modified phase-shifted CR force as the dominant correction to reacceleration rates (Equations \ref{eqn:phaseshift_stream} - \ref{eqn:streamGrowthMod_largeDiff}).

\begin{figure*}
\centering
\includegraphics[width=0.98\textwidth]{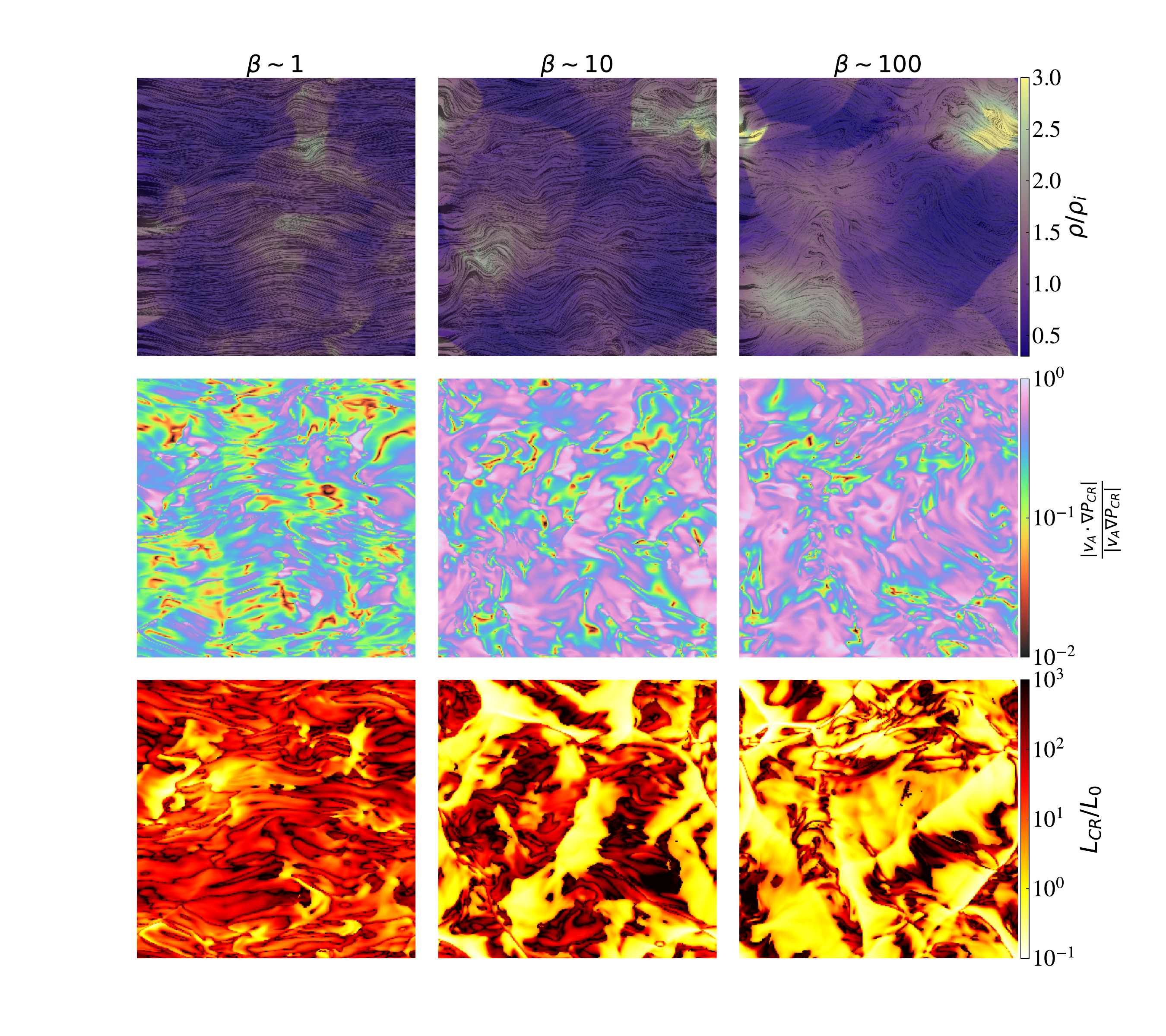}
\caption{Slice plots for three $256^{3}$ simulations, each with initial $P_{g}/P_{\rm CR} \sim 100$ but with $\beta \sim 1$, $M_{A} \sim 0.35$ (\emph{left}), $\beta \sim 10$, $M_{A} \sim 1.1$ (\emph{middle}), and $\beta \sim 100$, $M_{A} \sim 3.5$ . The top row shows gas density with magnetic field streamlines overlayed. The middle row shows the misalignment between {\bf B} and $\nabla P_{\rm CR}$. The bottom row shows the field-aligned CR scale length relative to the outer eddy scale. Notably, the higher $\beta$ simulations result in more field line tangling, generally greater alignment between {\bf B} and $\nabla P_{\rm CR}$  (leading to more CR energy loss), and shorter CR scale lengths.}
\label{fig:slicePlots}
\end{figure*}

Why does CR energy loss have this $M_{A}$ dependence? This largely arises due to misalignment between the magnetic field and CR pressure gradient. The streaming loss rate $\mathbf{v_A} \cdot \nabla P_{\rm CR}$ is sensitive to the angle between the magnetic field $\mathbf{B}$ and CR pressure gradient $\nabla P_{\rm CR}$. In $M_{A} < 1$ turbulence, magnetic tension is strong, and field line tangling is suppressed. While compressions (and hence $\nabla P_{\rm CR}$) can occur in all directions, the mean magnetic field maintains its initial orientation, so that $\mathbf{v_A} \cdot \nabla P_{\rm CR}$ is suppressed. This effect is apparent in Figure \ref{fig:slicePlots}, which shows slice plots of three $256^{3}$ simulations along the z = 0 axis after 5.8 eddy turnover times, when the turbulence is developed. Each column corresponds to different evolved plasma $\beta$, going from strong field (left) to weak field (right). The top row shows density with magnetic field lines overplotted; peaks and troughs are most apparent in the weak field case since magnetic tension is weakest, also allowing the field to tangle more easily. The middle row shows the collisionless energy loss rate, calculated as $|v_{A} \cdot \nabla P_{\rm CR}|$, divided by $|v_{A} \nabla P_{\rm CR}|$. This effectively quantifies the misalignment between the magnetic field and CR pressure gradient. Clearly, the $\beta \sim 1$ simulation shows large regions of misalignment, while the $\beta \sim 100$ simulation has a more isotropic magnetic field and, hence, better alignment between {\bf B} and $\nabla P_{\rm CR}$. 

Eventually, one would like a fitting formula for this loss rate for use in e.g. sub-grid models of CR transport for galaxy evolution simulations; however, that is premature at this stage. Clearly, the degree of misalignment depends on resolution at low $\beta$. For $\beta = 2$, our lowest resolution $64^{3}$ simulation has energy loss suppressed by a factor of $\sim 50$, while in our highest resolution $512^{3}$ simulation, the loss is suppressed by a factor of $\sim 10$. Higher resolution allows for a more accurate capture of field line tangling. Future higher resolution simulations are needed to probe convergence and facilitate the development of a sub-grid CR heating model, but we can clearly conclude that low-resolution galaxy evolution simulations, while possibly capturing reacceleration (which is dominated by the outer scale eddies) likely cannot capture true loss rates. However, at higher $\beta$, the suppression is weaker (approaching a few tenths) for all resolutions we tried, consistent with the field becoming more isotropic, and $t_{\rm CR, loss} \sim L_0/v_A$ is approximately realized.

An additional interesting effect is that streaming CRs create ``bottlenecks" on a timescale of order the Alfv\'{e}n crossing time. In the streaming dominated regime, $P_{\rm CR} \propto (v_A + v)^{-{\gamma_{CR}}}$ along a flux tube. A minimum in $(v_A + v)$ (e.g., due to a density spike) creates a situation where CRs would have to stream up their pressure gradient, which is not possible. Instead, CRs form a flat pressure profile where no momentum or energy are transferred to the gas \citep{Skilling1971, Wiener2017, Wiener2019, Bustard2021}; multiple regions can take the form of a `staircase' structure \citep{Tsung2021_staircase}. In sub-Alfv\'{e}nic turbulence, these bottlenecks, which form after an Alfv\'{e}n crossing time, can form before the flow randomizes during an eddy turnover time, potentially creating $\nabla P_{\rm CR} \approx 0$ regions where streaming energy loss is suppressed. We see this in the bottom row of Figure \ref{fig:slicePlots}, which shows the parallel (magnetic field-aligned) CR scale length $L_{\rm CR} = P_{CR}/\nabla_{||}P_{CR} = P_{CR}/(\hat{b} \cdot \nabla P_{CR})$ relative to the outer eddy scale. In large portions of the simulation box, $L_{\rm CR} >> L_{0}$ for $\beta \sim 1$. At higher $\beta$, due to the combined effects of greater field line tangling and slower development of bottlenecks, $L_{\rm CR} \sim L_{0}$ for much more of the volume. In 1D stratified media, bottlenecks have only weak effects on the {\it net} loss rate, though they greatly increase the spatial and temporal intermittency of energy loss \citep{Tsung2021_staircase}. In Bustard and Oh 2022, in prep, we probe CR energy loss / gas heating in greater detail and at higher spatial resolution, specifically how much gas heating occurs via grid-scale dissipation or via streaming energy loss. 

\begin{figure}
\centering
\includegraphics[width=0.46\textwidth]{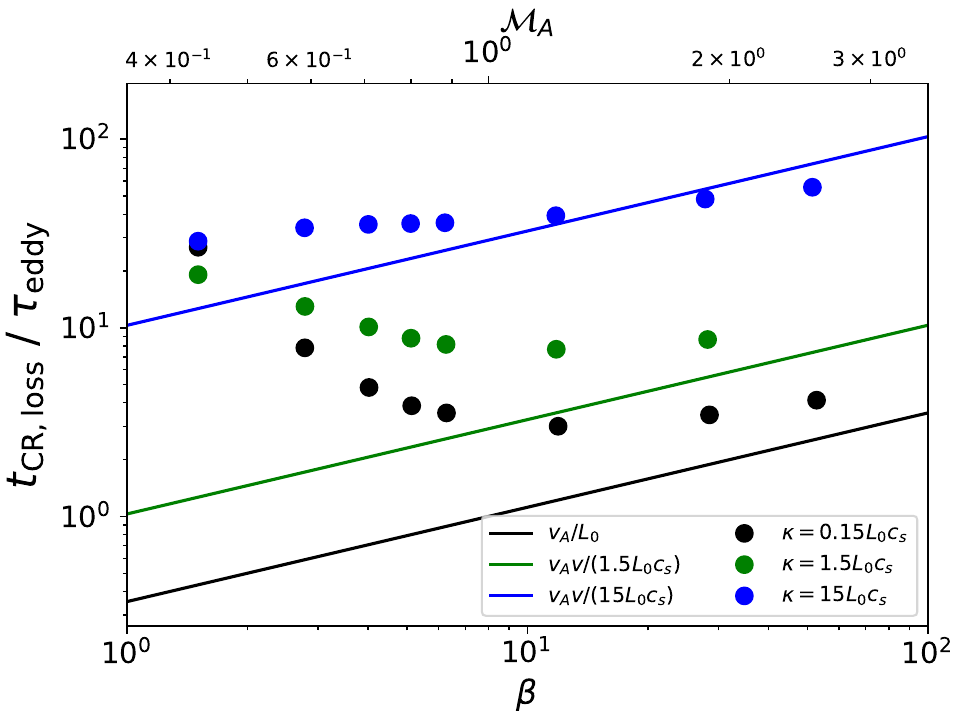}
\includegraphics[width=0.47\textwidth]{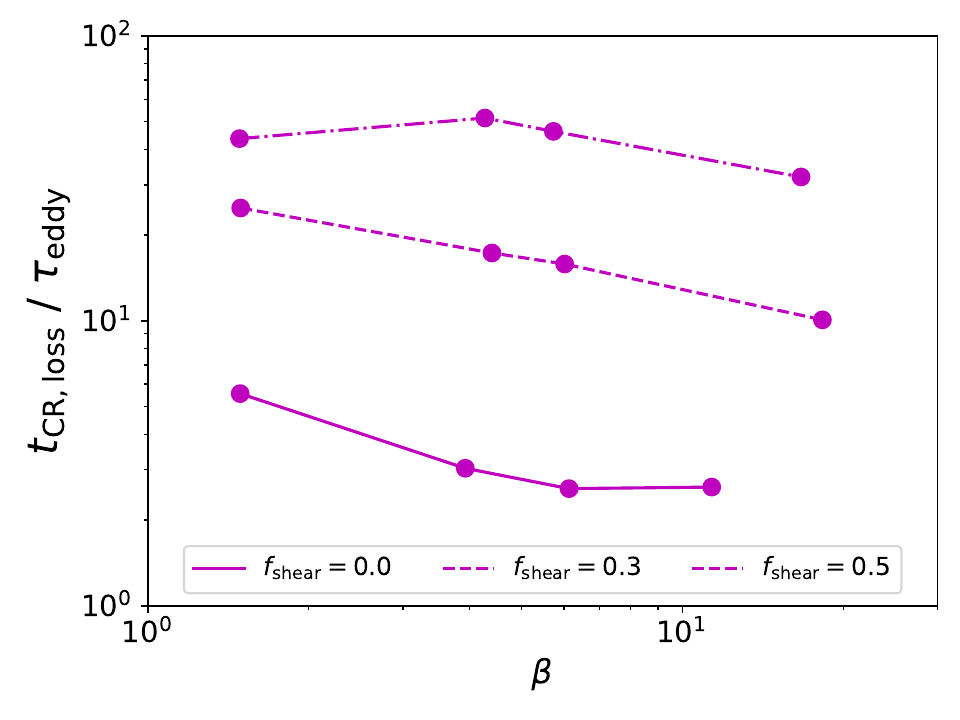}
\caption{\emph{Top:} Streaming energy loss time as a function of additional diffusivity; all simulations here were run on a $64^{3}$ grid. The solid lines now show $v_{A}/L_{\rm CR}$, where $L_{\rm CR} \sim \kappa / v$ when diffusion dominates and $L_{\rm CR} = L_{0}$ for the streaming-dominated simulation. At high $\beta$, these approximations are well reproduced. \emph{Bottom:} Streaming loss time as a function of $\beta$ for a mixture of solenoidal and compressive driving, controlled by the parameter $f_{\rm shear}$. All simulations were run on a $256^{3}$ grid. For increasingly solenoidal driving (increasing $f_{\rm shear}$), loss times and reacceleration times both increase.}
\label{fig:heating_varySol_varyKappa}
\end{figure}

\subsection{Varying the Diffusion Coefficient}
As a sanity check, we briefly explore sensitivities of the streaming energy loss time to varying $\kappa$, with each simulation run on a $64^{3}$ grid and with initial plasma $\beta$ varying from 2 to 200. As $\kappa$ increases, both the magnitude of energy loss and the discrepancy between simulated loss and expectation should decrease. In the top panel of Figure \ref{fig:heating_varySol_varyKappa}, the solid lines show the expected loss times, either $L_{0}/v_{A}$ for the streaming-dominated run or $(\kappa/ v)/v_{A}$ for the diffusion dominated runs, the difference owing to the appropriate guess for the CR scale height. Indeed, we see that, as the scale height increases with increasing diffusivity, the loss rate goes down for all $\beta$. Energy loss suppression due to misalignment of $\mathbf{B}$ and $\nabla P_{\rm CR}$ also decreases, since $\nabla P_{\rm CR}$ is no longer set by compression but by the speed of field aligned diffusion.

\subsection{Solenoidal vs Compressive Driving}
In hydrodynamic 3D turbulence, the fraction of power in compressive modes is given by \citep{Federrath2010}
\begin{equation}
    \frac{F_{long}}{F_{tot}} = \frac{(1-f_{\rm shear})^{2}}{1-2f_{\rm shear} + 3f_{\rm shear}^{2}}
\end{equation}
where $f_{\rm shear}$ is a stirring parameter we can vary between 0 (purely compressive forcing) and 1 (purely solenoidal forcing). A natural mixture of $f_{\rm shear} = 0.5$ yields $F_{long}/F_{tot} = 1/3$. Since CR pressure gradients, which lead to collisionless energy loss, are developed by compressive fluctuations, we expect the loss rate to be a declining function of $f_{\rm shear}$. The bottom panel of Figure \ref{fig:heating_varySol_varyKappa} shows a set of simulations, each on a $256^{3}$ grid and with initial plasma $\beta$ varying from 2 to 200, with the same total driving rate $\epsilon$ split into different mixtures of compressive and solenoidal driving. Note that, because solenoidal motions more easily amplify magnetic fields than compressive motions, the saturated plasma $\beta$ shown on the x-axis differs greatly from the initial plasma $\beta$, saturating, for example, near $\beta \sim 10$ ($\mathcal{M}_{A} \sim 1$) for initially super-Alfvenic turbulence. As expected, $f_{\rm shear} = 0$, corresponding to our purely compressive driving simulations shown in the top panel of Figure \ref{fig:heat_suppression}, gives the largest CR reacceleration rates ($\sim 72 t_{\rm eddy}$ for $f_{\rm shear} = 0$ vs $\sim 170 t_{\rm eddy}$ for $f_{\rm shear} = 0.5$) and the largest energy loss rates. Going from purely compressive forcing ($f_{\rm shear} = 0$) to a natural mixture ($f_{\rm shear} = 0.5$) decreases the energy loss rate by a factor of 5 or more. That this decrease is greater than a factor of 3, assuming $F_{\rm long}/F_{\rm tot} = 1/3$, can be understood if $\delta P_{CR}/P_{CR}$ and $\delta v/v$ are not well-coupled. We speculate that solenoidal motions lead to longer CR scale lengths due to the aforementioned bottleneck effect for streaming CRs; this decreases $\delta P_{CR}/P_{CR}$ and $\nabla P_{CR}$. This is worth studying in simulations where CRs are dynamically important, but for now, we just emphasize that solenoidal driving decreases both CR reacceleration rates and CR energy loss rates for a given driving rate $\epsilon$.

\section{Discussion}
\label{sec:discussion}

\subsection{Issues with Reacceleration in the ISM}

In propagation models, reacceleration is typically included as a default, and it is an attractive alternative to otherwise ``leaky box" models in that it provides a natural fit to low-energy ($R \approx 1$ GV) boron-to-carbon data while maintaining the standard paradigm of diffusive propagation, i.e. a single power-law dependence of the diffusion coefficient D(E) or, inversely, the escape path length $\lambda_{esc} \sim 1/D(E)$ \citep{Heinbach1995}. At the same time, canonical reacceleration rates are energetically troubling, as surprisingly high fractions of total CR energization (up to $\sim 50\%$, comparable to the contribution from diffusive shock acceleration) have been attributed to turbulent reacceleration \citep{Thornbury2014, Drury2017}, and they are increasingly in tension with data. If reacceleration were the dominant acceleration mechanism in the $\sim 1-100$ GeV range in our Galaxy, we would then see a progressive increase in secondary to primary ratio as a function of energy; instead, we see otherwise \citep{StrongReview2007, GabiciReview2019}. Additionally, at this low-energy end, a growing wealth of multiwavelength data, specifically synchrotron measurements, are best fit in models with little or no reacceleration \citep{Trotta2011, DiBernardo2013, Orlando2013, GabiciReview2019}, in tension with canonical reacceleration rates assuming purely diffusive CR transport. \emph{These low energy CRs, however, are precisely those which are self-confined, and where the impact of CR streaming must be considered.}

\cite{Hopkins2021_reacc} argues analytically that timescales for CR reacceleration, energy loss, and convection obey the ordering $\tau_{\rm conv} < \tau_{\rm loss} < \tau_{\rm reacc}$, in which case reacceleration is negligible and convection in a large-scale wind presents a compelling alternative to explain the bump in B/C at low energies. Our simulations specifically demonstrate that non-resonant reacceleration from subsonic, compressive turbulence is ineffective in typical ISM environments with low $\beta$, although here we find the suppression is due to phase-shifted CR forces rather than CR streaming losses. Streaming stunts energy gain due to large-scale compressive motions, even when CRs diffuse slowly and would otherwise gain energy from turbulence at the maximal rate. Even if typical ISM CR diffusivities lie in the $\kappa >> v_{ph} L_0$ regime\footnote{\label{footnote:variable_kappa} Note, however, that this presumes a constant diffusion coefficient. The empirical diffusion coefficients which are used in Galactic propagation models more likely reflect conditions in the halo, where diffusing particles spend most of their time. In quasi-linear theory of self-confinement, where diffusion expresses transport relative to the Alfv\'{e}n frame, the diffusion coefficient adapts to local conditions (e.g., see \citealt{Wiener2017}):
\begin{equation}
\frac{\kappa}{c_s L_0} \sim \beta^{-1/2} \left(\frac{L_c}{L_o} \right) \frac{\kappa}{v_A L_c} \sim  \beta^{-1/2} \left(\frac{L_c}{L_o} \right) \left( \frac{v_D}{v_A} -1 \right)
\label{eq:kappa_linear} 
\end{equation} 
where $v_{\rm D} \sim \mathcal{O}(v_{\rm A})$ is the net drift velocity. This can in fact result in ${\kappa}/{c_s L_0} \lsim 1$, rather than ${\kappa}/{c_s L_0} \gg 1$. }, where streaming corrections are smaller, reacceleration is, in any case, less efficient than in the well-trapped ($\kappa < v_{ph} L_0$) regime. That turbulence is not purely compressive further decreases the efficiency of non-resonant reacceleration since solenoidal motions don't energize CRs.
All considered, the neglect of streaming in Galactic transport codes (see \citealt{Hanasz2021} for a recent review) overestimates CR energization by second order Fermi acceleration by a large factor in $\beta \sim 1$ plasmas like the warm ISM.

What about other reacceleration mechanisms? Resonant second order Fermi acceleration, which relies on the presence of both forward and backward propagating waves, is not possible for self-confined CRs, for which the only waves that are excited are those that co-move with the CRs \citep{ZweibelReview2017}. Reacceleration by transit-time damping (TTD), essentially the magnetized version of Landau damping, is similarly inefficient at these energies: scattering rates from TTD are orders of magnitude lower than the rate of gyroresonant scattering by self-confinement (e.g. Figure 2 of \citealt{Yan2004}).

While we here consider only second order Fermi mechanisms, first order Fermi mechanisms, such as DSA or turbulent reconnection \citep{Lazarian2020}, may be efficient in regions of the ISM such as molecular clouds \citep{Gaches2021} or superbubbles \citep{Vieu2022} where turbulence is supersonic and therefore cascades into shocks. These first order processes imprint distinctly different spectral signatures than second order Fermi acceleration.

\subsection{Reacceleration in the CGM and ICM}    
While streaming considerably stunts reacceleration in low-$\beta$ galactic environments, non-resonant reacceleration in other environments is still quite plausible. Table \ref{table2} lists reasonable values for the phase velocity, outer driving scale, and plasma $\beta$ in the warm ISM (WIM), CGM, and ICM. 

The CGM is a plausible candidate for efficient non-resonant reacceleration. Collisional loss rates are long in these diffuse galaxy halos, and for typically assumed CR diffusion coefficients of $\kappa \sim$ few $\times 10^{28} \rm cm^{2}/s$, $\kappa < L_0 v_{ph}$ and CRs may be well-trapped in turbulent eddies of the CGM, prolonging their residence time in the CGM and boosting their energy density. The effect of streaming is tied to the local plasma $\beta$ and is therefore location-dependent. Observational constraints on plasma $\beta$ in the CGM are sparse, but observations of a fast radio burst passing through a foreground galaxy halo suggest $\beta > 1$ \citep{Prochaska2019}. In simulations, CGM plasma $\beta$ is often large (e.g., $\beta \sim 10-100$ in FIRE simulations; \citealt{Hopkins2019CRs}), though it can significantly fall in regions affected by galactic winds \citep{VanDeVoort2021}, with $\beta \sim 1$ most favorable in large-scale galactic winds. The efficacy of reacceleration could thus vary spatially, being most efficient in 
high-$\beta$ regions, while being stunted in others. Overall, turbulent reacceleration could play a much more significant role in regulating the CR content of the CGM, compared to the ISM.

\begin{table}
  \centering
  \caption{Typical values for the warm ISM (WIM), CGM, and ICM. Note that $\beta$ appears to vary quite widely in simulations of the CGM, e.g. \cite{VanDeVoort2021}, where $\beta \sim 0.01$ in localized regions coincident with galactic outflows, but $\beta \sim 10-100$ in quiescent regions. Recent observations of a fast radio burst passing through a foreground galaxy halo suggest $\beta > 1$ \citep{Prochaska2019}.}
  \begin{tabular}{l|c|c|c|}
    \toprule
    & WIM & CGM & ICM \\
    \hline
    $v_{ph}$ (cm/s) & $10^{6}$ & $10^{7}$ & $10^{7}$ \\
    $L_0$ & 100 pc & 1-10 kpc & 100 kpc \\ 
    $\kappa_{crit} = v_{ph} L_0$ (cm$^{2}$/s) & $3 \times 10^{26}$ & $3 \times 10^{28-29}$ & $3 \times 10^{30}$ \\
    $\beta$ & 1 & 1-10? & 100 \\
  \end{tabular}
\label{table2}
\end{table}

Reacceleration rates are largely unchanged by streaming losses in the high $\beta$ environment typical of the ICM, where reacceleration is frequently invoked to explain radio emission in merging galaxy clusters. CR energy densities in the ICM are constrained to be quite low but are still consistent with models of (efficient) turbulent reacceleration \citep{Brunetti2007, Brunetti2011}. While most theoretical analyses to-date of reacceleration in merging clusters have focused on reacceleration from TTD, we simply remind the reader that non-resonant reacceleration will, at some level, always be present and can have a competitive growth time compared to resonant reacceleration in high-$\beta$ environments (e.g., see Fig. 6 of \citealt{Brunetti2007}). This is useful to keep in mind due to the universality and simplicity of non-resonant reacceleration. The resonant case, on the other hand, relies on turbulence cascading down to gyroresonant scales and highly uncertain details of the spectra and damping scales in compressible MHD turbulence \citep{Brunetti2011,miniati15,Pinzke2017}. For instance, if compressible modes dissipate in weak shocks (Burgers turbulence), as is quite plausible, then particle acceleration rates are too low to explain observed giant radio halos. Even if compressible turbulence is Kraichnan, standard TTD on thermal particles gives problematic damping scales for turbulence, and a reduction of particle mean free path by plasma instabilities (mirror, firehose) might be necessary \citep{Brunetti2011}. By constrast, non-resonant turbulent reacceleration is a comparatively robust, well-understand mechanism further validated by our numerical simulations.

\subsection{Do Galaxy Evolution Simulations Accurately Capture CR Energy Transfer?}

In Appendix \ref{appendix:convergence}, we briefly review the impact of spatial resolution. We find that turbulent reacceleration is adequately resolved as long as the outer scale of turbulence is sufficiently well resolved (by $\sim 20$ cells). Most galaxy evolution simulations refine based on density, meaning they decrease spatial resolution going from the dense ISM to the diffuse halo. Fortunately, outer eddy scales similarly increase going from the ISM to the halo, meaning that simulations likely resolve at least the outer driving scale and, in fact, have a good chance of capturing accurate reacceleration rates. One caveat is turbulent reacceleration in the $\kappa \ll v_{\rm ph} L_0$ regime. As the dynamic range of the simulation increases, the acceleration time decreases, due to acceleration by smaller eddies, particularly those of size $l$ where $\kappa \sim v_{\rm ph} l$. 

Unfortunately, spatial resolution \emph{does} affect the influence of CRs on the background gas. When small-scale field line tangling is not well-resolved, streaming energy loss is artificially reduced (Figure \ref{fig:heat_suppression}), especially in our low-$\beta$ simulations. This has abundant implications for large-scale simulations. For instance, one effect of CRs that has garnered significant attention is their ability to drive multiphase galactic winds \citep{Ipavich1975, breitschwerdtwinds1991, SalemCROutflows2014, Ruszkowski2017, Buck2020, Hopkins2019CRs, Bustard2020, quataert21-streaming, Bustard2021, Huang2022}. Indeed, turbulent reacceleration could affect CR wind driving -- for instance, by re-energizing CRs in the halo despite strong energy losses in streaming scenarios arising from streaming down steep density gradients \citep{quataert21-streaming}. An unsolved issue with CR acceleration of multi-phase gas is how CRs accelerate cold ($T\sim 10^4$K) dense gas clouds. Do they provide direct acceleration by exerting CR forces on the cold gas, due to steep CR gradients which develop at the cold-hot gas interface \citep{Wiener2017,Wiener2019,Bruggen2020,Bustard2021}? Or is direct acceleration inefficient (as might be expected if field lines wrap around the cloud), and CRs first accelerate the background hot gas, which then accelerates the cold gas via mixing-induced momentum transfer \citep{Gronke2018,Gronke2020}? There are hints of the latter process in \citet{Bustard2021,Huang2022_arxiv}. The relative importance of indirect vs direct acceleration depends on the relative efficacy of CR thermal vs momentum driving, particularly in the background hot medium. 

Figure \ref{fig:heat_suppression} suggests that, for $M_{A} < 1$, the CR energy loss rate is a relatively flat or even rising function of $\beta$; a tangled field overcompensates for a lower magnetic field strength, at least until the field is isotropic ($M_{A} \sim 1$), at which point energy loss scales inversely with $\beta$. Hints of this behavior were recently seen in \cite{Huang2022_arxiv}, where CR heating and, in turn, indirect cloud acceleration were actually more efficient in runs with decreased field strength. A goal of future work, currently premature given the incomplete parameter space we've so far explored, should be to develop a sub-grid model for energy loss / gas heating as a function of resolution, $M_{A}$, and different turbulent driving modes. If convergence (not clearly seen in Figure \ref{fig:heat_suppression}) can be achieved with higher resolution runs, this may not be far off.

\section{Conclusions}
\label{sec:conclusions}

In this paper, we reviewed the non-resonant reacceleration of CRs in subsonic, compressive turbulence and derived heuristic modifications to reacceleration rates when CRs are self-confined or ``streaming." As CRs are believed to be self-confined up to 
$E \lsim 300$GeV CR reacceleration rates are modified throughout this entire energy range. 
After describing our analytic expectations, we ran a suite of MHD simulations to verify previously derived reacceleration rates for purely diffusive CRs \citep{Ptuskin1988, Chandran2004}, test the effects of streaming, and probe the nonlinear regime. Our main findings are as follows:

\begin{itemize}[left=4pt]
    \item Our simulations, which show a Burger's-like power spectrum, verify the analytic reacceleration rates derived for a $k^{-2}$ spectrum (\citealt{Ptuskin1988}; Equations \ref{eqn1},\ref{eqn2}), including the expected modifications due to anisotropic field-aligned transport at large $\kappa$ (\citealt{Chandran2004}; Equation \ref{eqn:anisotropic}). In particular, reacceleration rates peak for diffusion coefficients $\kappa \sim 0.1 L_0 v_{\rm ph}$, and have scalings consistent with analytic expectations. To our knowledge, this is the first time CR turbulent reacceleration has been shown in MHD simulations with a fluid CR treatment and is an encouraging test of the Athena++ CR module implemented in \cite{JiangCRModule}. 
    
    \item When CR streaming is introduced, the rate of net energy gain can be substantially suppressed compared to the diffusion only case, due to modified phase shifts between CR and gas variables. For $\kappa \lsim L_{0} v_{ph}$, the regime where reacceleration is canonically most efficient, growth times in low-$\beta$ plasmas typical of the ISM are significantly longer than canonical expectations (Equation \ref{eqn:streamGrowthMod}; right panel of Fig \ref{fig:CR_Energy_betaStudy}). By contrast, in high $\beta$ environments like the CGM and ICM, streaming does not have a significant impact on turbulent reacceleration. In diffusion-dominated regimes ($\kappa > L_{0} v_{ph}$), the impact of streaming is milder, but reacceleration is in any case already inefficient. New reacceleration times $t_{\rm grow} \sim \frac{p^{2}}{D_{pp}}$ in $\kappa < v_{ph} L_{0}$ and $\kappa > v_{ph} L_{0}$ regimes, respectively, are given below, assuming a $k^{-2}$ kinetic energy spectrum (see Equation \ref{eq:Dpp_Wk} and the ensuing discussion for how to calculate reacceleration rates with alternative power spectra):
    
    \begin{align}
    t_{\rm grow} \sim & \frac{9}{2} \frac{v_{\rm ph} L_0}{v^{2}} f_{\rm corr} \left({\rm tan^{-1}}\left(\frac{\kappa}{v_{\rm ph}L_1}\right)\right.  \\ 
    & - \left. \rm tan^{-1} \left(\frac{\kappa}{v_{\rm ph}L_0}\right) \right)^{-1} \nonumber
    \label{eqn:final_tgrow2}
    \end{align}
    $L_{0}$ is the outer eddy scale, $L_{1}$ is the smaller characteristic scale of shocks in the medium, and $f_{\rm corr}$ encodes corrections due to anisotropic diffusion \citep{Chandran2004} and streaming transport assuming an isothermal gas (this work): 
    \begin{align}
        f_{\rm corr} = & \frac{1}{1 - \sqrt{2/ \beta}} \quad (\kappa < v_{ph}L_0)  \\
         = & \left( \frac{v L_{\rm turb}}{\kappa_{\parallel}} \right)^{1/2} \left(\frac{1}{1 - \sqrt{2/ \beta}}\right)^{1/2} \quad (\kappa \gg v_{ph}L_0) 
         \label{eq:f_corr1}
    \end{align}
    The $\beta$-dependent terms are relevant for self-confined CRs with energy $E \lessapprox 300$ GeV; at higher energies, CRs are no longer self-confined and these can be dropped.

\end{itemize}    
To diagnose the limitations of lower-resolution galaxy evolution simulations and as a step towards sub-grid modeling of CR energetics and influence, we also determine some sensitivities to resolution.

\begin{itemize}[left=4pt]
    \item Reacceleration rates with pure diffusion are largely insensitive to resolution (the minimum growth time is well-captured even when the outer eddy scale is only resolved by 20 cells), but higher resolution more accurately captures power at small scales, boosting simulated reacceleration rates. This is important in the $\kappa \ll v_{\rm ph} L_0$ regime.
    
    \item Streaming energy loss $v_{A} \cdot \nabla P_{CR}$ is more strongly dependent on Alfv\'{e}n Mach number $M_{A}$ and simulation resolution. Because streaming energy loss only occurs when the CR pressure gradient is aligned with the magnetic field, not every compression induces streaming energy loss. This misalignment between the magnetic field and $\nabla P_{\rm CR}$ is a function of plasma $\beta$ since it is more difficult for turbulence to tangle magnetic field lines in highly magnetized plasmas. The net result is that relative CR loss rates are much lower than $v_{A}/L_{0}$ in low-$\beta$ plasmas (Figures \ref{fig:heat_suppression} and \ref{fig:slicePlots}) and are clearly sensitive to resolution. When field line tangling is resolved, the average misalignment between $\nabla P_{\rm CR}$ and the magnetic field decreases. Counterintuitively, due to the countervailing effects of increased CR streaming speeds and decreased field alignment at lower $\beta$, our highest resolution results over $\beta \sim 1-100$ show that CR energy loss is insensitive to plasma $\beta$. This also shows that the reduced acceleration rates at low $\beta$ is not due to increased CR losses.
    
\end{itemize}

An important issue we have not considered in this paper is the effect of density stratification, which results in a background CR gradient. This provides constant CR coupling, pressure support and heating, and if sufficiently strong, can drive a wind. How does CR reacceleration proceed in such a background? One issue is that the density fluctuations created by turbulence can create CR `bottlenecks' \citep{Skilling1971,Wiener2017}: small decreases in the Alfv\'{e}n speed along a magnetic flux tube cause CRs to pile up or ``bottleneck." For purely streaming CRs, they readjust to these conditions and create a flat CR pressure upstream of the dip in Alfv\'{e}n speed. As $\nabla P_{\rm CR} \rightarrow 0$, CRs no longer excite confining Alfv\'{e}n waves and instead free-stream at close to the speed of light, no longer transferring energy or momentum to the gas. It is as yet unclear how this stochastic coupling affects CR energization and escape rates from stratified, turbulent media. We have also not considered the case where CR phase shifts and/or heating conspire to reverse the sign of energy transfer, such that CRs give energy to gas motions, rather than vice-versa \citep{Begelman1994,Tsung2021_staircase}. This would occur when $\beta \le 0.25$, a regime we plan to study in future work.  

Finally, although it is clear that CRs can damp compressive turbulence, their back-reaction on the turbulent cascade is relatively unexplored but possibly significant, for instance, in CR-dominated galaxy halos \citep{JiCRHalos2020}. We explore this in a series of higher-resolution follow-up simulations (Bustard and Oh 2022, in prep), where we find that CRs, even with very low reacceleration rates, can absorb a significant fraction of large-scale turbulent energy, subsequently modifying the compressive cascade.

\section*{acknowledgments}

The authors gratefully acknowledge Navin Tsung, Max Gronke, Yan-Fei Jiang, Christoph Federrath, Hui Li, and Ellen Zweibel, as well as the organizers and participants of the KITP ``Fundamentals of Gaseous Halos" workshop. CB was supported by the National Science Foundation under Grant No. NSF PHY-1748958 and by the Gordon and Betty Moore Foundation through Grant No. GBMF7392. SPO was supported by NSF grant AST-1911198, and NASA grant 19-ATP19-0205.

Computations were performed on the Stampede2 supercomputer under allocations TG-PHY210004, TG-AST190019, and TG-AST180036 provided by the Extreme Science and Engineering Discovery Environment (XSEDE), which is supported by National Science Foundation grant number ACI-1548562 \citep{xsede}.

\software{Athena++ \citep{AthenaRef}, yt \citep{ytPaper}, Matplotlib \citep{matplotlib}, Mathematica \citep{Mathematica}}

\appendix

\section{Convergence of Reacceleration Rates} 
\label{appendix:convergence}

\begin{figure}
\centering
\includegraphics[width=0.68\textwidth]{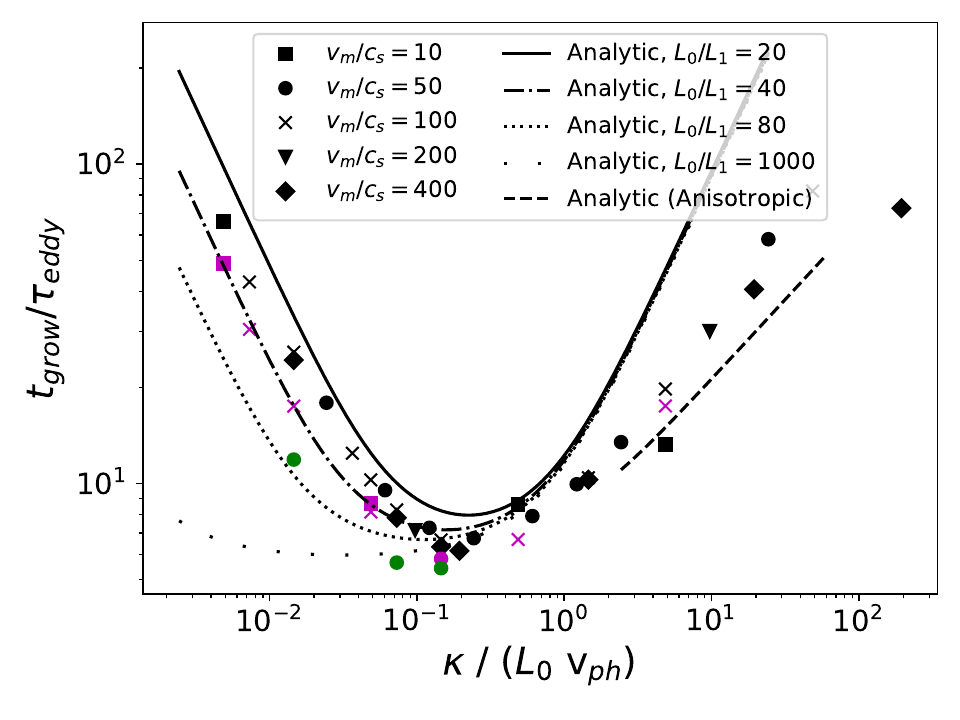}
\caption{Testing growth time convergence with respect to the reduced speed of light $v_m$ (relative to the fastest propagation speed, $c_{s}$), and numerical resolution. Simulations with $v_{m}/c_{s} = 50$ appear well-converged. Magenta and green points denote simulations with $128^{3}$ and $256^{3}$  resolution, respectively, for which the velocity divergence and hence compressional heating is slightly larger and decreases the reacceleration time in the well-trapped $\kappa < v_{ph} L_{0}$ regime. For large $\kappa$, CRs diffuse quickly over small-scale structures, which then do not need to be well-resolved to get converged reacceleration rates.}
\label{fig:appendixFig}
\end{figure}

We now study simulation convergence, with respect to both spatial resolution and choice of the maximum speed of light in the two-moment method. The time-dependent CR flux term $1/v_{m}^{2} \partial F_{\rm CR}/\partial t$ in the two-moment equations allows CRs to free-stream at the speed of light when the CR pressure gradient vanishes: $\nabla P_{\rm CR} \rightarrow 0$ and $\sigma_{c} \rightarrow 0$. If the flux term is sufficiently small, CRs are well-coupled to waves, and the two-moment equations of CR hydrodynamics collapse to the usual one-moment equations \citep{breitschwerdtwinds1991}. To accurately describe coupling vs decoupling, some care must be taken with the flux term, specifically the value of $v_{m}$, which must be greater than all other speeds in the system (including the CR propagation speed). We fiducially set $v_{m} = 50c_{s}$, which gives converged reacceleration rates at not only our fiducial resolution of $2L/\Delta x = 64$ but also at higher resolutions $2L/\Delta x = 128, 256$. While not shown here, we also find that $v_{m} = 50c_{s}$ is necessary to get converged CR loss rates, with lower $v_{m}$ artificially boosting loss rates. Figure \ref{fig:appendixFig} shows simulation data with maximum speed of light in the range $v_{m}/c_{s} =$ 10 to 400, each for purely compressive turbulence with turbulent Mach number $\mathcal{M} \sim 0.5$. Simulated growth times are well converged with respect to $v_{m}$ and show an excellent match to the analytic derivation from \cite{Ptuskin1988}, assuming $L_{0}/L_{1} = 20$.  

This convergence may seem a bit surprising at first since the time-dependent flux term $1/v_{m}^{2} \partial F_{\rm CR}/\partial t$ is only small when $v_A/v_{\rm m} < \Delta x/L$; however, this is only the criterion for the two-moment equations to effectively collapse to the one-moment equations. Many of the test problems in e.g. \cite{JiangCRModule, Navin2021} violate this criteria but show convergence to analytic results. Indeed, one of the most powerful components of the two-moment method is the ability to unlock stability and convergence from the quadratic time-stepping requirement needed for one-moment implementations.

Convergence with respect to spatial resolution proves a bit trickier in our simulations. For instance, in Figure \ref{fig:CR_Energy_betaStudy}, one can see that the growth times for pure diffusion continue to decrease with increasing resolution. The value of $L_1$, which we associate with the width of a shock front in the medium, should be set by the spatial resolution. For our $64^{3}$ simulations, the outer scale is resolved by $\sim 20$ cells in each direction, making $L_0/L_1 \sim 20$. In real astrophysical plasmas, though, the scale separation is much larger, and the velocity divergence, which ultimately energizes CRs, gets additional contributions from smaller scales. Lines in Figure \ref{fig:appendixFig} denote different $L_0/L_1$. Notably, the growth times in the $\kappa < v_{ph} L_{0}$ regime are much shorter; presumably, simulations of higher resolution would match these curves. To test this, we ran a few $128^{3}$ and $256^{3}$ simulations, again using $v_{m}/c_{s} = 50$. Growth times in the $\kappa < v_{ph} L_{0}$ regime do decrease: The $128^{3}$ runs match the analytic curve with $L_0/L_1 = 40$ quite well; the $256^{3}$ runs, for which we have a smaller set of data points, show somewhat shorter growth times than the minimum of the $L_0/L_1 = 80$ curve, but the growth time near $\kappa/v_{ph} L_{0} \sim 0.01$ returns to the $L_0/L_1 = 80$ curve. Overall, we recover the expected trend that higher resolution in the $\kappa < v_{ph} L_{0}$ regime should lead to shorter growth times. Since CRs with $\kappa > v_{ph} L_{0}$ diffuse quickly over small-scale structures, higher resolution makes very little difference in this regime.

\bibliographystyle{apj}
\bibliography{bibliography, peng_citations}

\end{document}